\documentclass[aps,pre,twocolumn,showpacs,amsmath,amsfonts,
eqsecnum,superscriptaddress]{revtex4} \usepackage{epsfig}
\usepackage{bm}

\newcommand{\Onecol} {\begin{widetext} \onecolumngrid} %% 2 -> 1
\newcommand{\Twocol} {\end{widetext} \twocolumngrid}   %% 1 -> 2
\newcommand{\Ref}[1]{(\ref{eq:#1})}%%  requires \Ref{label}
\newcommand{\REF}[1]{Eq.~(\ref{eq:#1})}%%  requires \Ref{label}
\def\etal{{{ et~al.}}} \def\eg{{{e.g.}}}
\def\etc{{{{etc.}}}}   
\def\vs{{{vs.~}}} \renewcommand{\it}[1]{\textit{#1}}
\renewcommand{\S}[2]{\section{\label{s:#1}{#2}}}
\renewcommand{\SS}[2]{\subsection{\label{ss:#1}{#2}}}
\newcommand{\SSS}[2]{\subsubsection{\label{sss:#1}{#2}}}
\newcommand{\BE}[1]{\begin{equation}\label{eq:#1}}    %% Equation
\newcommand{\EE}{\end{equation}}%%   environment with label eq:#1
\newcommand{\BEA}[1]{\begin{eqnarray} \label{eq:#1}} %%  The same
\newcommand{\EEA}{\end{eqnarray}} %%           for Equation Array
\newcommand{\BEa}{\begin{eqnarray*}} %%              Not numbered
\newcommand{\EEa}{\end{eqnarray*}}   %%            Equation Array
\def\nn {\nonumber} \def\br {\\ \nonumber} %% Breaks line in Eqs.
\newcommand{\B}[1]{{\bm{#1}}}%% Bold Roman & Greek Lower & Upper Case
\newcommand{\C}[1]{{\mathcal{#1}}}    %%   Calligrapfic Upper case
\newcommand{\BC}[1]{\bm{\mathcal{#1}}}%% Bold Calligrapfic Upper case
\def\k{\B {k}} \def\r{\B {r}} %% Bold k  &  r
\def\d{\text{d}} \def\i{\text{i}}%% Roman  d & i
\def\l{\ell} \newcommand{\ve}{\varepsilon} %% \ell & \varepsilon
\newcommand{\p}{\partial}           %% \partial
\def\<{\left \langle} \def\>{\right\rangle}%% Adjustable < & >
\def\Re{{\C R}\mkern-3.1mu e}  %%         Reynolds number  Re
\def\Rec{{\C R}\mkern-3.1mu e_{\text{cr}}} %%  Re_{cr}
\newcommand{\ok}{(\omega,\B k)}
\newcommand{\oke}[1] {(\omega_{#1},\B k_{#1})} %% with extension #1
\newcommand{\tr}{(t,\B r)} 
\newcommand{\tk}{(t,\B k)}

\newcommand{\pot}[1]{\frac{\partial {#1}}{\partial t}}

\newcommand{\Int}[2]{\int\frac{\d^{#1}{#2}}{(2\,\pi)^{#1}}}

\renewcommand{\sb}[1]{_{\text {#1}}}  %% sub-   for lower case
\renewcommand{\sp}[1]{^{\text {#1}}}  %% super- for lower case
\def\Fbox#1{\vskip1ex\hbox to 8.5cm{\hfil\fboxsep0.3cm\fbox{%
  \parbox{8.0cm}{#1}}\hfil}\vskip1ex\noindent}  %%  {TEXT} in BOX
\def\K41{\fboxrule0.2ex\fbox{\large\text{K41}}}%% Big K41 in BOX

\begin{document}
\title{Effect of particle inertia on the turbulence in a suspension}
%%%%%%%%%%%%%%%%%%%%%%%%%%%%%%%%%%%
\author{Victor S. \surname  L'vov}
\email{Victor.Lvov@Weizmann.ac.il}
\homepage{http://lvov.weizmann.ac.il}
%\thanks{Thanks to all people around the word for their help}
\affiliation{Department of Chemical Physics, The Weizmann Institute
of Science, Rehovot 76100, Israel}
%%%%%%%%%%%%%%%%%%%%%%%%%%%%%%%%%%
\author{Gijs Ooms}
\email{G.Ooms@wbmt.tudelft.nl}
\affiliation{J.M. Burgerscentrum, Laboratory for Aero- and
Hydrodynamics,  Technological University Delft, Mekelweg 2, 2628 CD
Delft, The Netherlands}
%%%%%%%%%%%%%%%%%%%%%%%%%%%%%%%%%%
\author{Anna Pomyalov}
\email{Anna.Pomyalov@Weizmann.ac.il}
\affiliation{Department of Chemical Physics, The Weizmann Institute of
Science, Rehovot 76100, Israel}
%%%%%%%%%%%%%%%%%%%%%%%%%%%%%%%%%%%%%%%%%%%%%%%
\begin{abstract}
 We propose a  \it{one-fluid} analytical model for  a turbulently
flowing dilute suspension, based on a modified Navier-Stokes
equation with a $k$-dependent effective density of suspension,
$\rho\sb {eff}(k)$, and an additional damping term $\propto
\gamma\sb p(k)$, representing the fluid-particle friction
(described by Stokes law). The statistical description of
turbulence within the model is simplified by a modification of the
usual closure procedure based on the Richardson-Kolmogorov picture
of turbulence with a differential approximation for the energy
transfer term. The resulting ordinary differential equation for
the energy budget is solved analytically for various important
limiting cases and numerically in the general case. In the
inertial interval of scales we describe analytically two competing
effects: the energy suppression due to the fluid particle friction
and the energy enhancement during the cascade process due to
decrease of the effective density of the small scale motions. An
additional suppression or enhancement of the energy density may
occur in the viscous subrange, caused by the variation of the
extent of the inertial interval due to the combined effect of the
fluid-particle friction and the decrease of the kinematic
viscosity of the suspensions. The analytical description of the
complicated interplay of these effects supported by  numerical
calculations is presented. Our findings allow one to rationalize
the qualitative picture of the isotropic homogeneous turbulence of
dilute suspensions as observed in direct numerical simulations.
\end{abstract}


%\date{Version of \today}
\maketitle \Onecol\large
%%%%%%%%%%%%%%%%%%%%%%%%%%%%%%%%%%%%%%%%%%%%%%
\section*{\label{s:intro} Introduction
} The interaction of solid particles or liquid droplets with the
turbulence in a gas controls the performance of various
engineering devices and is important for many practical
applications like the combustion of pulverized coal and liquid
sprays and cyclone separation. This interaction plays also an
important role in many areas of environmental science and physics
of the atmosphere. Dust storms, rain triggering, dusting and
spraying for agricultural or forestry purposes, preparation and
processing of aerosols are typical examples. For a review of
turbulent flows with particles and droplets see, \eg~the book by
C.T. Crowe, M. Sommerfeld and Y.Tsuji~\cite{B2}.

In dilute suspensions with small volume fractions of particles,
$C\sb p$, the particle-particle interactions are negligible.
Nevertheless, for $\rho\sb p/\rho\sb f\gg 1$ (the ratio of the
solid particle material and the gas densities), the mass loading
$\phi= C\sb p \rho\sb p/ \rho\sb f$ may exceed unity and the
kinetic energies of the particles and the carrier gas may be
comarable. Hence the ``two-way coupling'' effect of the fluid on
the  particles and vice versa must be accounted for. Current
understanding of the turbulence in dilute suspensions is still at
its infancy due to the highly nonlinear nature of the physically
relevant interactions and a  wide spectrum of governing parameters
(the particle size $a$ \vs $L$ and $\eta$, the outer and inner
scales of turbulence, the particle response time $\tau\sb p$ \vs
$\gamma_{_L}$ and $\gamma_\eta$, the turnover frequencies of $L$-
and $\eta$- scale eddies).

Existing analytical studies of the problem are mainly based upon a
\it{two-fluid} model description  wherein both the carrying fluid
and particle phases are treated as interpenetrating
continua~\cite{B2,83EA,99DE,98BSS}. This model deals with
non-interacting solid spherical particles with a radius $a$ small
enough such that:
\begin{enumerate}
\item One can neglect the effect of preferential concentration and
may assume homogeneity of the particle space distribution. This is
not always so. Above some critical radius $a\sb{cr}$ the space
homogeneous distribution of particles becomes unstable. Resulting
clustering instability leads to  preferential concentration.  For
a detailed theory of this effect, see Ref.~\cite{BeerSheva} and
references therein. In the present paper we consider only
particles with $a<a\sb{cr}$.

\item The Stokes viscous drag law for particle acceleration, $d \B
u\sb p/d t= [\B u\sb f-\B u\sb p]/\tau\sb p$, is valid ($\B u\sb
f$ is the fluid velocity).

\end{enumerate}

 Unfortunately, the statistical description of two-fluid
turbulence with closure procedures requires a set of additional
questionable simplifications due to the lack of understanding of
the relevant physics of the particle-fluid interactions. This
makes  closures of the two-fluid model highly qualitative at
best~\cite{94Egl,98BSS,00BSS}.

We think that the basic physics of the problem may be better
described by a simpler \it{one-fluid model} for turbulent dilute
suspensions, which uses standard closure relations of one-phase
turbulence. The present paper suggests such a model and, as a
first step, uses a properly modified simple closure, based on the
Kolmogorov-Richardson cascade picture of turbulence.  The
resulting non-linear differential equation for the energy budget
were solved analytically. This provides an economical and
internally consistent analytical description of the turbulence
modification by particles including the dependence of suppression
or enhancement of the turbulence on the three governing
parameters: $(\tau\sb p \gamma_{_L})$, $\phi$ and the scale of
eddies. These effects were previously observed in numerous
experimental and numerical publications, see, \eg the review by
Crowe, Trout and Chung~\cite{B3}. Many groups carried out
experimental work; for an overview see Pietryga \cite{01Pie}.
Other researchers studied the modification of turbulence by small
particles using direct numerical simulations
\cite{90SE,93ET,98BSS,99SC,01Dru} or by large-eddy simulation
\cite{B13}.  Nevertheless  the complicated physics of turbulently
flowing suspensions in the two-way coupling regime still wait for
a detailed  analytical description.

Our analytical findings in this paper successfully correlate
important features of turbulence modification observed in
numerical simulations Refs.~\cite{98BSS,01Dru,99SC}. We believe
that the one-fluid model (together with more advanced closures of
one-phase turbulence) offers an insight in  basic physics of
particle-laden turbulent flows. The next step in this development
should include the effect of preferential concentrations, which
was studied so far only for a given turbulent flow field of the
carrier fluid~\cite{BeerSheva}.

The paper is organized as follows.  In Sec.~\ref{s:NSE} we review,
after a presentation of the notation (Sec.~\ref{ss:notations}) and
an evaluation of the characteristic time scales
(Sec.~\ref{ss:timescales}), some publications about
DNS-simulations (Sec.~\ref{ss:DNS}), about experimental work
(Sec.~\ref{ss:lab-exp}) and about some analytical
models(Sec.~\ref{ss:models}). A critical evaluation of the
existing analytical models\cite{98BSS,71BP,77AT,B5,B6,B7,B11} is
made.

In Sec.~\ref{s:basic} we suggest a new \it{one-fluid} equation of
motion~\Ref{NS-basic} for turbulently flowing suspensions with
small particles. This is a modified version of the
Navier-Stokes equation with two wave-number-dependent parameters,
$\rho\sb{eff}(k)$ and $\gamma\sb p(k)$:
\begin{itemize}
\item The $k$-dependent effective density of suspensions
$\rho\sb{eff}(k)$ describes the different degree of involvement of
heavy particles in turbulent fluctuations with different wave-numbers
[referred to below as $k$-eddies]. For $k$-eddies with a turnover time
$1/\gamma(k)$, which is much smaller than the particle response time
$\tau\sb p$, the particles may be considered at rest and
$\rho\sb{eff}(k)$ is about the density of the fluid itself, $\rho\sb
f$. For $k$-eddies with $\tau\sb p \gamma(k)\ll 1$ the effect of the
particle inertia may be neglected and particles may be considered as
fully involved in the motion of eddies.  Therefore for small enough
$k$ the effective density $\rho\sb{eff}(k)$ is close to the mean
density of the suspension (fluid plus particles), $\rho\sb s=\rho\sb
f(1+\phi)$.  Our
\REF{r-eff} reasonably describes $\rho\sb{eff}(k)$ for all values
of $k$.

\item The damping term $\gamma\sb p(k)$, given by \REF{gamma-p}
describes the  fluid-particle viscous friction. The function
$\gamma\sb p(k)$  saturates at the level  $ 1/\tau\sb p$ for small
scale eddies with $\tau\sb p \gamma(k)\gg 1$, when the particles
may be considered to be almost at rest. In this regime the damping
is $k$-independent, while the turnover frequency of $k$-eddies
$\gamma(k)$ grows with $k$. Therefore for large $k$ $\gamma\sb
p(k)\ll \gamma(k)$ and the particle-induced damping of these
$k$-eddies may be neglected  with respect to  their energy loss in
the cascade process, which is determined by the frequency
$\gamma(k)$.  In contrast, for small enough $k$ [when $\tau\sb p
\gamma(k)\ll 1$] the particles are \it{almost completely} involved
in the motions of $k$-eddies and their contribution to $\gamma\sb
p(k)$ is suppressed by the factor $[\tau\sb p \gamma(k)]^2 \ll 1$
with respect to $ 1/\tau\sb p$.
\end{itemize}

Our one-fluid model for  turbulent suspensions~\Ref{NS-basic} is
first postulated in Sec.~\ref{ss:adv}.  Its physical
interpretation is discussed in Sec.~\ref{ss:NS-physics}. A
detailed derivation of~\REF{NS-basic} is given in
Secs.~\ref{ss:ass}, \ref{ss:NSo-derivation}
and~\ref{ss:NSk-derivation}. The most difficult problem here is
how to account for the nonlinear effect of the interaction of
$k$-eddies within the \it{one-fluid} model of turbulent
suspensions. The suggested form of the nonlinear
term~\Ref{NL-basic} is a modification of the standard
Navier-Stokes nonlinearity and is based on:
\begin{itemize}
\item a rigorous description of eddy interactions in both limiting
cases $\tau\sb p\gamma(k)\ll 1$ and $\tau\sb p\gamma(k)\gg 1$
\item  respect of the fundamental symmetries of the problem --
Galilean invariance and conservation of energy.
\end{itemize}

Section~\ref{s:balance} deals with the budget of the kinetic
energy in turbulently flowing suspensions. One has to account not
only for the dissipation of energy due to the fluid-particle
friction but also for the effect of particles on the energy
redistribution in the system due to the eddy interaction. First we
derive in \ref{ss:energy} the budget equation~\Ref{budget} which
accounts for the energy pumping due to a stirring  force, energy
damping due to the kinematic viscosity and fluid-particle friction
and also describes the flux of energy over the scales due to the
nonlinearity of the problem. Equation~\Ref{budget} is exact but
unfortunately is not closed. As usual it includes a 3rd order
velocity correlation functions. As a first step in the analysis of
turbulent suspensions in the framework of our one-fluid
model~\REF{NS-basic} and the budget~\REF{budget}, we use in this
paper, sect.~\ref{ss:closure}, a simple closure procedure based on
the Richardson-Kolmogorov cascade picture of turbulence in which
the energy flux is accounted for in a differential approximation.
Needless to say that there are various closure procedures for the
Navier-Stokes turbulence in the literature. They may be
straightforwardly applied to our \REF{NS-basic}. This important
part of the project will be done elsewhere.

 The derived energy balance equations are summarized in
Sec.~\ref{ss:basic-eq}. They have a very simple and transparent
analytical form~\Ref{bud2} -- \Ref{rel-1}, allowing their
effective analytical analysis, see Sects.~\ref{s:prelim} and
\ref{s:modification}. In particular in Sec.~\ref{ss:no-part} we
found a simple solution for the case of micro-particles having a very
small response time.  In Sec.~\ref{ss:energy-bud} we found the
iterative solution for the case of a suspension with heavy particles
in the inertial interval and analyzed its accuracy in
Sec.~\ref{ss:accuracy}.

In section~\ref{s:modification} we analytically describe a
complicated interplay between two competitive effects: of the
turbulence suppression and the turbulence enhancement in the
inertial interval of scales, as well as in the viscous subrange. A
brief comparison of our finding with DNS results is done in
Sec.~\ref{ss:comp}.

In the concluding Sec.~~\ref{s:sum} we summarize the results of the
paper and present our ideas for further work.
\section{Notations and relevant timescales}
\label{s:notaions}

%
%%%%%%%%%%%%%%%%%%%%%%%%%%%%%%%%%%%%%%%%%%%%%%%
\SS{notations}{Nomenclature}

\begin{itemize}

\item  $\rho\sb f$,  $\B u\tr $ -- density and velocity of the fluid

\item $\tilde {\B u}(\omega, \B r)$, ${\B u}\tk $, $\tilde {\B u} \ok
$ -- Fourier transform of $\B u\tr$ with respect to $t$ [$t\to
\omega$], with respect to $\B r$ [$\B r\to \k$], and to both variables
[$t\to \omega\,,\ r\to \k$]

\item $F\tk$, $\tilde{F}\ok$ -- pair correlation functions of fluid
velocity in $\tk$ and $\ok$ representation

\item $E(k)=\rho\sb f \, k^2 F(0,k)/2 \pi $ -- one dimensional
spectrum of the turbulent kinetic energy of the pure fluid (fluid
without particles)

\item $\C E(k)$ -- one dimensional spectrum (of the turbulent
kinetic energy) of the suspension

\item $\gamma(k)$ -- turnover frequency of $k$-eddies (turbulent
fluctuations of the characteristic scale $1/k$). May be understood
also as $1/\tau(k)$, where $\tau(k)$ is the life time of
$k$-eddies. In the Kolmogorov~41 picture of turbulence
$\gamma(k)\simeq k \sqrt{k E(k)/\rho\sb f}$.

\item $E=\int\d k E(k)/2\pi$, \ $\C E=\int\d k \C E(k)/2\pi$ -- total
turbulent kinetic energy of respectively the pure fluid and the
suspension

\item  $a$, $\rho\sb p$, $m\sb p=4\pi a^3 \rho\sb p /3 $ --
radius, density and mass of the particles

\item   $C\sb p$, $\l^3=1/C\sb p $ , volume fraction of particles
and volume of suspension per particle

\item $\psi = [4\pi a^3/3]/\l^3$,  $\phi=m\sb p/ \rho\sb f \l^3$
-- volume fraction and  mass loading parameter

\item $\tau\sb p$ -- particle response time, also referred to as
\it{Stokes time scale}

\item $ \tau_{_L}$ -- turnover time   of the energy containing
eddies (of scale $L$)

\item $\delta\equiv \tau\sb p / \tau_{_L}$ -- the particle
response time in the units of  $ \tau_{_L}$.

\item $\eta$; $v_\eta$, $\tau_\eta=\eta/u_\eta$ -- Kolmogorov
(viscous) microscale; characteristic velocity and time at
scale~$\eta$ of turbulence \item $\rho\sb{eff}(k)$ -- effective
density of the suspension for turbulent fluctuations of
characteristic scale $1/k$ [referred to as $k$-eddies]

\item $\nu$, $\nu\sb{eff}(k)$ -- kinematic viscosity of the pure
fluid, effective kinematic viscosity of $k$-eddies in the suspension

\item $\gamma\sb p (k)$ -- effective damping frequency in the
suspension due to the fluid-particle friction

\item $\varepsilon(k)$ -- (one dimensional) flux of the turbulent
kinetic energy of the suspension via a sphere of radius $k$ in
$\k$-space, also referred to as \it{energy flux over scales}.
\end{itemize}
%%%%%%%%%%%%%%%%%%%%%%%%%%%%%%%%%%%%%%%%%%%%%%%%%%%%%%%%%%%%

\SS{timescales}{Evaluation of time scales}

The radius of the particles is supposed to be small enough, so that
the particle Reynolds number $\Re\sb p$ is less than a critical value
($\Rec$). In this case we can apply the Stokes approximation
(according to which the fluid-particle friction force is proportional
to the difference between the particle velocity and the fluid
velocity). Careful analysis by Lumley ~\cite{B10} shows that in a
turbulent flow the condition for the validity of  $\Re\sb
p\alt \Rec$ may be expressed via the particle radius $a$ and the
Kolmogorov micro-scale $\eta$ in the following way
\begin{eqnarray}
  \label{eq:St-apll}
   a \alt 2 \eta(\rho\sb f/ \rho \sb p)^{1/3}.
\end{eqnarray}

It is clear that one of the important parameters  in the physics of
turbulently flowing suspensions is the ratio of the inertial time
scale of the particles (the Stokes time scale $\tau\sb p$) and the
life time $\tau_\eta$ of eddies of the Kolmogorov micro-scale. The
particle response time is given by
\begin{eqnarray} \label{eq:Stime}
\tau\sb p &=& \frac{m\sb p }{ 6 \pi \,\nu\, \rho \sb f \, a } =
\frac{2\, \rho\sb p \, a^2 }{9\,  \rho\sb  f\,  \nu }\,,
\end{eqnarray}
where we use the  expression for the particle mass $m\sb p$:
\begin{eqnarray}
  \label{eq:mass1}
  m\sb p =\frac{4 \pi}{3} a^{3}\rho\sb p\ .
\end{eqnarray}
As is well-known the Kolmogorov micro-scale $\eta$ is found from the
condition that the Reynolds number for eddies of scale $\eta$ is equal
to unity:
 \begin{eqnarray}
   \label{eq:ReK}
   \Re_{\eta}= \eta \, v_\eta/ \nu =1\ .
 \end{eqnarray}
Here $ v_\eta$ is the characteristic velocity of $\eta$-scale
eddies. It is related to the turnover time of these eddies in the
following manner $\tau_\eta=\eta/v_\eta$.  This allows us to rewrite
the requirement~\Ref{ReK} as follows
\begin{eqnarray}\label{eq:Ktime}
  \tau_\eta =\eta^2/\nu\ .
\end{eqnarray}
The ratio of the time-scales $\tau\sb p$ and $\tau_\eta $ immediately
follows from Eqs.~\Ref{Stime} and \Ref{Ktime}:
\begin{eqnarray}\label{eq:ratiotimes1}
\frac{\tau\sb p} {\tau_ \eta}=\frac{2}{9}\,\frac{\rho\sb
p}{\rho\sb f}\,\frac{a^2}{\eta^2}\ .
\end{eqnarray}
Substituting the condition~\Ref{St-apll} for the validity of the
Stokes approximation we find
\begin{eqnarray}\label{eq:t-rat}
\frac{\tau\sb p} {\tau_\eta}\alt
\left(\frac{\rho\sb p}{\rho\sb f}\right)^{1/3}\,,
\end{eqnarray}
where we neglected the difference between $8/9$ and 1.
Equation~\Ref{t-rat} means that for ``heavy'' particles in a gas,
that satisfy Stokes approximation, the particle response time
scale may be about ten times larger than the Kolmogorov time
scale: $\tau\sb p \alt 10 \tau_\eta$.  For such particles in a
liquid the two time scales are about the same. So we may conclude
that for ``heavy'' particles in a gas, that satisfy Stokes
approximation, the inertia of the particles may be expected to be
important in a considerable part of the energy spectrum. For
particles in a liquid the particle inertia will only be
significant for the smallest eddies, for which $\tau\sb p \simeq
\tau_\eta$.
%%%%%%%%%%%%%%%%%%%%%%%%%%%%%%%%%%%%%%%%%%%%%%%%%%%%%%%%%%%%%%%%%
\S{NSE}{ Review of literature }
%%%%%%%%%%%%%%%%%%%%%%%%%%%%%%%%%%%%%%%%%%%%%%%%%%%%%%%%%%%
 This section is devoted to a review of the literature about the
 problem of a turbulently flowing suspension. We will review
 important findings
 from published numerical experiments, physical experiments and
 analytical models.
%%%%%%%%%%%%%%%%%%%%%%%%%%%%%%%%
\SS{DNS}{Review of some DNS simulations}

To study the two-way coupling effect several groups have applied the
direct-numerical-simulation technique (DNS) to particle-laden
isotropic turbulence. A brief review of some of the publications is
given below.

Squires and Eaton~\cite{90SE} considered the particle motion in
the Stokes regime in which gravitational settling was neglected.
They assumed statistically stationary isotropic turbulence. Mass
loadings from zero to unity were considered for a series of
particle response times varying from $0.3\,\tau_\eta$ to $11\,
\tau_\eta$, where $\tau_\eta$ is the Kolmogorov time scale. They
found that the overall reduction in turbulence kinetic energy for
increasing mass loading was insensitive to the particle response
time. They attributed the non-uniform distortion of the turbulent
energy spectrum by particles to the preferential concentration of
particles into regions of low vorticity and/or high strain rate.

~Elghobashi and Truesdell~\cite{93ET} examined turbulence
modulation by particles in decaying isotropic turbulence. They
used the particle equation of motion derived by Maxey and
Riley~\cite{83MR}, and found that for the large density ratio
considered in their simulations the particle motion was influenced
mostly by drag and gravity. They found that the coupling between
particles and fluid resulted in an increase in small-scale energy.
This increase in the energy of the high-wave-number components of
the velocity field resulted in a larger dissipation rate. They
also found that the effect of gravity resulted in an anisotropic
modulation of the turbulence and an enhancement of turbulence
energy levels in the direction aligned with gravity.

Boivin, Simonin and Squires~\cite{98BSS} also made a very detailed
DNS-study of the modulation of isotropic turbulence by particles.  The
focus of their work was on the class of dilute flows in which particle
volume fractions and inter-particle collisions are
negligible. Gravitational settling was also neglected and the particle
motion was assumed to be governed by drag with particle response times
ranging from the Kolmogorov scale to the Eulerian time scale of the
turbulence and particle mass loadings up to unity. The velocity field
was made statistically stationary by forcing the low wave-numbers of
the flow. Like in ~\cite{90SE,93ET} the effect of particles on the
turbulence was included by using the point-force approximation. The
DNS-results showed that particles increasingly dissipate fluid kinetic
energy with increased mass loading, with the reduction in kinetic
energy being relatively independent of the particle response time (as
was already found in~\cite{90SE}). The viscous dissipation in the
fluid decreases with increased mass loading and is larger for
particles with smaller response times. The fluid energy spectra show
that there is a non-uniform distortion of the turbulence spectrum with
a relative increase in small-scale energy (as was found
in~\cite{93ET}). They state that the fluid drags the particles at low
wave-numbers while the converse is true at high wave-numbers for small
particles.

Sundaram and Collins~\cite{99SC} performed DNS-simulations of
particle-laden isotropic decaying turbulence. The particle
response time was in the range $1.6\ \tau_{\eta} \alt \tau_{p}
\alt 6.4\ \tau_{\eta}$. The ratio of the particle density and
fluid density was of the order $10^3$. The particle Reynolds
number $\Re\sb p$ remained less than $\Re \sb {cr}$, and the drag
force on the particles was described by Stokes law. The
point-force approximation was employed to represent the two-way
coupling force in the fluid momentum equation. The DNS-results
showed that the particles reduce the turbulent kinetic energy as
compared to the particle-free case, and this reduction is less
pronounced for smaller response times $\tau_{p}$. The results also
showed that the total turbulent energy dissipation is increased by
the particles, and the increase is larger for smaller $\tau_{p}$.
The turbulent energy spectrum is reduced at small wave-numbers and
increased at high wave-numbers by the two-way coupling, and the
location of the cross-over point is shifted towards larger
wave-numbers for larger $\tau_{p}$.

Druzhinin ~\cite{01Dru} examined the modulation of isotropic
decaying turbulence by microparticles, for which $2a<\eta$,
$\tau\sb p < \tau_\eta$ and $\Re \sb p<\Re \sb {cr}$.  The
gravitational settling is neglected. Due to the fact that $\rho\sb
p \gg \rho\sb f$, the mass loading may be large enough to modify
the carrier flow. Druzhinin first derived an approximate
analytical solution for the energy spectrum and then performed
also DNS simulations. The results obtained for particles whose
$\tau\sb p \le 0.4 \tau_\eta $ show that both the turbulence
kinetic energy and the turbulence dissipation rate are increased
by the two-way coupling effect as compared to the particle-free
case. For particles with sufficiently high inertia ($\tau\sb p \ge
0.5 \tau_\eta $) the two-way coupling effect caused a reduction in
the turbulence kinetic energy as compared to the particle free
case. Druzhinin, therefore, showed that there occurs a qualitative
transition in the two-way coupling effect of particles on
isotropic turbulence as the particle response time is increased
from $\tau\sb p \ll \tau_\eta $ (microparticles) to $\tau\sb p
\simeq \tau_\eta $ (particles with finite inertia). For
microparticles there is an increase of all wave-numbers in the
energy spectrum. For particles with a higher inertia that is no
longer the case.

\SS{lab-exp}{Review of some laboratory experiments}

Many experiments have been carried out to study the modulation of
turbulence in the carrier phase by particles. An overview of the
experimental work up to 1999 is given by
Pietryga~\cite{01Pie}. Experimental measurements in shear flows,
\eg~particle-laden jets and boundary layers, have shown that
turbulence velocity fluctuations may be either increased or decreased
due to the modulation of the flow by (heavy) particles. However in
turbulent shear flows it is often difficult to separate the direct
modulation of the turbulence due to the momentum exchange with the
particles from the indirect changes occurring through modification of
turbulence production mechanisms via interaction with mean
gradients. In grid-generated turbulence these production mechanisms
are absent. It approximates in the best possible way the homogeneous,
isotropic turbulence with particles that we study in this
publication. We will, therefore, briefly review below some literature
publications about experimental work devoted to the study of the
modulation by particles of grid-generated turbulence.

Schreck and Kleis~\cite{93SK} studied the effect of almost
neutrally buoyant plastic particles (density 1050 kg/m$^3$) and
heavy glass particles (density 2400 kg/m$^3$) on grid-generated
turbulence in a water flow facility. The average particle size was
$655 \mu$m. The particle Reynolds number of the plastic particles
$\Re\sb p\approx 8$, for the glass particles $\Re\sb p\approx 20$.
The particle volume fraction was varied between $ 0.4 \%$ and $1.5
\%$, so the system was very dilute. Mean velocity and velocity
fluctuations of both phases were measured by a laser-Doppler
velocimeter. The presence of the particles in sufficiently high
concentration modified the turbulence downstream of the grid. The
decay rate of the turbulence energy increased monotonically with
particle concentration. The additional dissipation rate for the
suspensions with the heavier glass particles was about double that
of the almost neutrally buoyant plastic particles. A simple model
based on the slip velocity between the phases under-predicted the
measured increase in the dissipation rate. Schreck and Kleis,
therefore, assumed that a large portion of the additional
dissipation is associated with the measured modification of the
spectral distribution of the turbulence energy. They speculate
that the particles enhance the transfer of energy to smaller
eddies extending the dissipation spectrum to smaller scale. Since
only part of the high wave-number end of the spectrum could be
resolved experimentally, this speculation could not be
conclusively demonstrated by their experimental data.

Hussainov \etal~\cite{00HKR} studied the modulation of grid-generated
turbulence by coarse glass particles in a vertical downward channel
flow of air. Two different types of grids were used. Glass beads with
an average diameter of $700 \mu$m and a mass loading of $10 \%$ were
used. The particles were about 7 times larger than the Kolmogorov
length scale $\eta$ and $\Re\sb p\approx 70$ or 93, dependent on the
type of grid used. The particle response time scale of the particles
$\tau\sb p$ was about 5000 to 7000 times larger than the Kolmogorov
time scale $\tau_\eta$. The mean velociy and the turbulence intensity
along the channel axis (and in some cross-sections) were measured by
means of a laser-Doppler velocimeter. The decay curves of the
turbulence intensity in the streamwise direction showed an attenuation
of turbulence intensity of the flow by the particles. The particles
caused an increase in the total dissipation rate of the
turbulence. Hussainov \etal~found that the presence of the particles
decreased the energy spectra at high frequencies. This seems to be in
contradiction with the speculation of Schreck and Kleis, that the
particles enhance the transfer of energy to smaller eddies.
%%%%%%%%%%%%%%%%%%%%%%%%%%%%%%%%%%%%%%%%%%%%%%%%%%%%
\SS{models}{Analytical models}

The starting point for analytical models, described in the literature,
is often the Navier-Stokes (NS) equation for the velocity of the pure
fluid (fluid without particles) $\B u\tr$
\begin{eqnarray}\label{eq:NSf}
\rho\sb f \Big[ \pot {} + (\B u \cdot \B \nabla) - \nu \nabla^2
\Big]\B u + \B \nabla p = \B f\sb p+ \B f \,,
\end{eqnarray}
where $p\tr$ is the pressure and $\rho \sb f $ is the fluid
density. The random vector field $\B f\tr$ represents the stirring
force responsible for the maintenance of the turbulent
flow. Equation~\Ref{NSf} includes also the force $ \B f\sb p\tr $
caused by the friction of the fluid with particles:
\begin{eqnarray}
  \label{eq:friction}
  \B f\sb p \tr = \frac{\phi \rho\sb f }{\tau\sb p} \big[\B v \tr-\B
  u\tr \big]\ .
\end{eqnarray}
Here $\B v \tr$ is the velocity field of the particles, considered as
a continuous medium with density $m\sb p/\l^3=\rho\sb f \phi$, where
$m\sb p$ is the mass of a particle, $\l^3$- suspension volume per
particle and $\phi$ is the mass loading parameter
\begin{eqnarray}\label{eq:loading}
\phi= m\sb p / \rho \sb f\, \ell^3 \ .
\end{eqnarray}
The validity to represent $\B f\sb p \tr$ in the form of
~\Ref{friction} is based on the assumption of space homogeneity of
the particle distribution. It is also assumed that the particles
are small enough for the Stokes drag law to be valid. The equation
of motion, suggested in the literature, for the continuum phase of
the particles does often not include the pressure and viscous
terms
\begin{eqnarray}
  \label{eq:NSp}
  \Big(\frac{m\sb p}{\l^3}\Big) \Big[ \pot {} + (\B v \cdot \B \nabla)
  \Big]\B v = - \B f\sb p\ .
\end{eqnarray}

Equations~\Ref{NSf} and~\Ref{NSp} were used by Baw and
Peskin~\cite{71BP} to derive a set of ``energy balance'' equations for
the following functions:
\begin{itemize}
\item $E\sb{ff}(k)$ -- energy spectrum of the fluid turbulence, $E(k)$
in our nomenclature
\item $E\sb{ff,p}(k)$ -- energy spectrum of the fluid turbulence along
a particle trajectory
\item $E\sb{fp}(k)$ -- fluid-particle covariance spectrum
\item $E\sb{pp}(k)$ -- particle energy spectrum
\end{itemize}
In the balance equations the following energy transfer functions occur
\begin{itemize}
\item$T\sb{ff,f}(k)$ -- energy transfer in fluid turbulence
\item$T\sb{fp,f}(k)$ -- transfer of fluid-particle correlated motion
by the fluid turbulence along the particle path
\item$T\sb{fp,p}(k)$ -- transfer of fluid-particle correlated motion by
the particles
\item$T\sb{pp}(k)$ -- transfer of particle-particle correlated motion
by the particle motion
\end{itemize}
\begin{itemize}
\item $\Pi\sb{q,f}(k)$ -- fluid-particle energy exchange rate.
\end{itemize}
Baw and Peskin~\cite{71BP} made a set of simplifying assumptions in
order to be able to analyze the balance equations. First, they assumed
that the particles do not respond to the fluid velocity fluctuations
due to their (very large) inertia. Therefore
\begin{equation}\label{eq:BP1}
  \hskip -2.5cm E\sb{ff,p}(k)=E\sb{ff}(k)\,, \
\end{equation}\vskip -0.8cm
\[
T\sb{fp,f}(k)=T\sb{fp,p}(k)=T_{pp,p}(k)=0\ .
\]
This assumption is, of course, not realistic for particles satisfying
the Stokes' approximation. Their next assumption
\begin{equation}\label{eq:BP2}
\Pi\sb{q,f}=\phi[E\sb{fp}(k)-E\sb{ff,p}(k)]/ \tau\sb p\,,
\end{equation}
may be understood as a statement that the fluid-particle exchange rate
is statistically the same for all scales characterized by a
$k$-independent frequency $\gamma\sb p=\phi/\tau\sb p $. This is
reasonable for particles with very large inertia, but then Stokes law
is not valid. For particles satisfying Stokes law, assumption
~\Ref{BP2} has to be replaced with a more realistic, $k$-dependent
frequency $\gamma\sb p(k)$. We will come back to this point when
discussing our new model.

A serious difficulty in the derivation of the energy balance equations
is how to find a closure expression for third-order velocity
correlation functions, responsible for the various energy transfer
functions. Baw and Peskin assumed that $T\sb{ff,f}(k)$ can be
expressed similarly as in the case of a pure (single phase) flow
\begin{equation}\label{eq:BP3}
T\sb{ff,f}(k)=-\frac{\d}{\d\, k} \frac{
\epsilon_{f}^{1/3}k^{5/3}E\sb{ff}(k) }{\alpha} \,,
\end{equation}
where $\epsilon\sb f$ is the viscous dissipation in the pure fluid
(without particles) and $\alpha$ is the so-called Kolmogorov
constant. This assumption seems questionable to us. According to the
spirit of the Richardson-Kolmogorov cascade picture of turbulence one
may express inertial range objects, like $T\sb{ff,f}(k)$ in terms of
\it{again} inertial range quantities, like $k$, $E\sb{ff}(k)$ (which
is done in ~\REF{BP3}) and $\ve(k)$, the energy flux in $k$-space. In
a single phase flow, indeed $\ve(k)=\epsilon\sb f$. However this is not
the case for a turbulent suspension due to the fluid-particle energy
exchange, given by ~\REF{BP2}. We think that our closure (to be
discussed later on) is an improvement in this respect.

With this simplified model Baw and Peskin predicted the following
influences on the energy spectrum of the fluid turbulence due to the
particles:
\begin{itemize}
\item a decrease of the energy in the energy-containing range of the
spectrum
\item an increase in the inertial range of the spectrum
\item a decrease in the viscous dissipation range.
\end{itemize}

Boivin, Simonin and Squires~\cite{98BSS} used the same model as in
Ref.~\onlinecite{71BP}. They also applied assumptions similar to
~\REF{BP2} and ~\REF{BP3}. Fortunately, they took into account the
response of the particles to the turbulent velocity fluctuations by
relaxing assumptions~\Ref{BP1} and also accounted for the very
important physical effect of the energy dissipation due to the drag
around the particles. For that reason they approximated
$T\sb{ff,f}(k)$ and $T\sb{fp,f}(k)$ as follows:
\begin{equation}\tag{BSS1}
T\sb{ff,f}(k)=-\frac{\d}{\alpha\,\d k}
\big[ \epsilon\sb f-\Pi\sb{q,f}(k)\big]^{1/3}k^{5/3}E\sb{ff}(k) \,,
\end{equation}
\vskip -0.4cm\[
\hskip -0.8cm T \sb{fp,f}(k)=-\frac{\d}{\alpha\,\d k}
\big[\epsilon\sb f-\Pi\sb{q,f}(k)\big]^{1/3}k^{5/3}E\sb{fp}(k)\ .
\]
Notice that this closure has the same weakness as ~\REF{BP3},
involving the dissipation range value $\epsilon\sb f$. With the above
described changes with respect to the model described in
Ref.~\onlinecite{71BP} Boivin, Simonin and Squires found an increase
in the viscous dissipation range of the fluid turbulence spectrum for
small values of the particle response time $\tau\sb p$.

Al Taweel ~\cite{77AT} calculated the rate of additional energy
dissipation due to the presence of the particles. Because of their
inertia the particles were assumed not to follow completely the
turbulent velocity fluctuations of the carrier fluid. They expressed
the additional dissipation in terms of the turbulent kinetic energy of
the suspension. Then they added this term to the balance equation of
the turbulent kinetic energy, making the (questionable) assumption
that the energy flux across the spectrum has the same functional form
as in a single-phase flow. Solving this equation they found an
attenuation of the high-frequency fluctuations with a small alteration
of the energy-containing low frequencies. Although there was an
additional energy dissipation due to the particles, the total energy
dissipation was reduced due to the reduction of viscous dissipation in
the carrier fluid.

In a number of publications~\cite{B5,B6,B7,B11} Felderhof, Ooms
and Jansen developed an analytical model for the dynamics of a
suspension of solid spherical particles in an incompressible fluid
based on the linearized version of the Navier-Stokes equation. In
particular they studied the effect of the particles-fluid
interaction on the effective transport coefficients and on the
turbulent energy spectrum of the suspension. Also the hydrodynamic
interaction between the particles and the influence of the finite
size of the particles were incorporated. However it is needless to
say that the nonlinearity of the Navier-Stokes equation is of
crucial importance in the problem of turbulence. Felderhof, Ooms
and Jansen were well aware of this problem, but wanted to study in
particular the influences of the particle-particle hydrodynamic
interaction and of the finite particle size at a high particle
volume concentration.

\S{basic}{One-fluid model Navier-Stokes Equation for
 turbulent suspensions}

In Sec.~\ref{ss:models} we discussed the \it{two-fluid} model of
suspensions consisting of the Navier-Stokes (NS) equation~\Ref{NSf}
for the fluid and Eq.~\Ref{NSp} for the "gaseous" phase of
particles. This approximation is based on the assumptions of space
homogeneity of the particle distribution and applicability of the
Stokes drag law for the fluid-particle friction.  We think that the
basic physics of a turbulently flowing suspension with these
assumptions may be described in the framework of the much more simple
\it{one-fluid} equation. This model is presented in Sec.~\ref{ss:adv},
discussed in Sec.~\ref{ss:NS-physics} and ``derived'' in
Secs.~\ref{ss:NSo-derivation} and
\ref{ss:NSk-derivation}.

\SS{adv}{The model} The following equation may be considered as a
model equation for  turbulently flowing suspensions:
\begin{eqnarray}
  \label{eq:NS-basic}
&& \rho\sb{eff}(k) \big [\pot{}+
 \gamma\sb p (k) + \gamma_0 (k)] \B u\tk \br
&=&  - \BC N \{\B u,\B u \}_{t,\k} + \B f\tk \,,
\end{eqnarray}
The linear part of this equation involves:
\begin{eqnarray}
  \label{eq:r-eff}
&&\hskip -1.6cm \rho\sb{eff}(k)=\rho \sb f\Bigg\{1-  \psi+\phi\,
\frac{1+ 2 \tau\sb p
\gamma(k) }{\big[1+ \tau\sb p \gamma(k)\big]^2}\Bigg\} \,,\\
\label{eq:gamma-p} &&\hskip -1.5cm  \gamma\sb
p(k)=\frac{\phi\,\tau\sb p[\gamma(k)]^2}{ (1+ \phi)[1+ 2 \tau\sb p
\gamma(k)] + [\tau\sb p \gamma(k)]^2}\,,
\\ \label{eq:nu-eff}
 &&  \hskip -1.5cm \gamma_0 (k) =  \nu \sb{eff}(k)k^2\,, \quad
\nu\sb{eff}(k) =\frac{\nu\, \rho\sb f}{\rho\sb{eff}(k)} \ .
\end{eqnarray}
The nonlinear term in \REF{NS-basic} has the usual  NS equation
form:
\begin{eqnarray} \nn
  \C N\{\B u,\B u \}_{t,\k}^\alpha &=&\int\frac{\d^3 k_1\, \d^3 k_2 }
 {(2\pi)^3} \Gamma^{\alpha\beta\gamma}_{\B k\,\B k_1\B
  k_2} u_\beta^*(t,\k_1) u_\gamma^* (t,\k_2)
 \ . \\&&  \label{eq:NL-basic}
\end{eqnarray}
However the vertex $\Gamma^{\alpha\beta\gamma}_{\B k\,\B k_1\B
k_2}$ differs from the standard vertex
$\gamma^{\alpha\beta\gamma}_{\B k\,\B k_1\B k_2}$ of the NS
equation (see \eg ~Refs.~\onlinecite{B8,B4}):
\begin{eqnarray}   \nn
 \gamma^{\alpha\beta\gamma}_{\B k\,\B k_1\B k_2}&=&\frac{\rho\sb f}{2}
\big[P^{\alpha\beta}(\B k)\,k^\gamma +P^{\alpha\gamma}(\B k)\,
k^\beta\big] \delta(\B k+\B k_1+\B k_2) \,, \\ \label{eq:vertex}
\end{eqnarray}
as follows:
\begin{eqnarray}
    \Gamma^{\alpha\beta\gamma}_{\B k\,\B k_1\B k_2}= \rho\sb
    {eff}\Big(\frac{2\,k_1k_2k_3}{k_1^2+k_2^2+k_3^2}\Big)
    \frac{\gamma^{\alpha\beta\gamma}_{\B k\,\B k_1\B k_2}}{\rho\sb f}\ .
\label{eq:vertex-basic}
\end{eqnarray}

Our model  differs from the standard NS equation in three aspects:

\it a.  Equation \REF{NS-basic} involves the $k$-dependent
effective density of suspensions $\rho\sb {eff}(k) $ given by
\REF{r-eff}. The function $\rho\sb {eff}(k)$ satisfies the
inequality $\rho\sb f \le\rho\sb {eff}(k)\le \rho\sb f (1+\phi)$.
One could say that $\rho\sb {eff}(k) - \rho\sb f$ represents the
contribution of the particles involved in turbulent fluctuations
with characteristic scale $1/k$ to the effective density of
suspensions.

\it b. Equation \Ref{NS-basic}  includes the additional damping
term $\gamma\sb p(k) $,~\REF{gamma-p}, describing the loss of
kinetic energy caused by the viscous fluid-particle friction.

\it c.  In the absence of a stirring force $\B f\tr$ and both
damping terms, \REF{NS-basic} conserves the total kinetic energy
of suspensions $\C E$ [given by \REF{Eks2}] which is different
from the kinetic energy $E$ of the fluid itself.

The explicit form (\ref{eq:NL-basic}) of the nonlinear term is not
necessary for the simple closure procedure that we applied in this
publication. For the introduction of the energy flux in used
closure procedure it is enough to use the fact that the modelled
nonlinearity must be conservative. However, the explicit form is
needed for more advanced closure procedures that we intend to use
in future work. For this reason we  include it in this
publication.

%%%%%%%%%%%%%%%%%%%%%%%%%%%%%%%%%%%%%%%%%%%%%%%%%%%%%%%%%%%%%%
\SS{NS-physics}{Physical interpretation of the one-fluid model}

In a simplified fashion we may interpret $\rho\sb {eff}(k)$, the
$k$-dependent density of suspension in our model
equation~\Ref{NS-basic} as follows.

Denote as $f\sb {com}(k)$ the fraction of particles \it{co-moving}
with the $k$-eddies (turbulent fluctuation with some wave-number
$k$), in the sense that their velocity is the same as the velocity
of $k$-eddies. These particles also participate in the motion of
eddies with smaller wave-number $k'<k$ but not necessarily in the
motion of $k''$-eddies with $k''>k$. For small $k$ the turnover
frequency $\gamma(k)$ of $k$-eddies is small in the sense
$\gamma(k)\tau\sb p\ll 1$. Therefore, in this region of $k$, the
particle velocity is very close to that of the carrier fluid and
we can describe the suspension as \it{a single fluid} with
effective density $\rho\sb {eff}(k)$, which is close to the
density of suspension:
\begin{eqnarray}\label{eq:sus-den}
\rho\sb s&=& \rho\sb f(1-\psi)+ C\sb p \rho\sb p=\rho\sb
f(1-\psi+\phi)
\,,\br %%%%%%%%%%%%%%%%%
 \psi&\equiv& C\sb  p [4\pi a^3/3]\,,
 \quad \phi=C\sb p\,\rho\sb p/ \rho\sb f\ .
\end{eqnarray}
Here $C\sb p$ is the particle concentration, $\psi$ and $\phi$ are
the volume fraction and mass loading parameter.  However, for
large $k$, when $\gamma(k)\tau\sb p\gg 1$, the particles cannot
follow these very fast motions and may be considered at rest. Thus
the particles do not contribute to the effective density and
$\rho\sb {eff}(k)\to \rho\sb f$. In the general case $\rho\sb
{eff}(k)$ may be written as
\begin{eqnarray}
  \label{eq:eff-den}
   \rho\sb {eff}(k)=\rho\sb f \big[1-\psi + \phi f\sb{com}(k)]\,,\quad
\end{eqnarray}
Here a statistical ensemble  of all particles, partially involved
in the motion of $k$-eddies, is replaced by two sub-ensembles of
``fully co-moving'' (fraction $f\sb{com}(k)$ ) and ``fully at
rest'' (fraction $ f\sb{rest}(k)=1-f\sb{com}(k)$)  particles,
which does not contribute to $\rho\sb{eff}(k)$.

 The particles at rest cause the fluid-particle friction. According to
 Newton's third law, the damping frequency of a suspension $\gamma\sb
 p(k)$ may be related to the particle response time, $\tau\sb p$, via
 the ratio of total mass of particles $M\sb p$ at rest to the total effective
 mass of the suspension $M\sb{eff}(k)$:
\begin{eqnarray}
\label{eq:gamma-p1}
\gamma\sb p(k)=\frac{M\sb p}{\tau\sb p \, M\sb{eff}(k)}
= \frac{ C\sb p\, \rho\sb p
f\sb{rest}(k)}{\tau\sb p\, \rho\sb {eff}(k)}
= \frac{\phi\,  \rho\sb f \, f\sb{rest}(k)\,
}{ \tau\sb p\, \rho\sb {eff}(k)}\ .
\end{eqnarray}
As we mentioned, the fractions $f\sb{com}(k)$ and $f\sb{rest}(k)$
depend on $\tau\sb p\gamma(k)$. Moreover, the portion
$f\sb{rest}(k)$ is independent on the sign of the velocity,
therefore we expect $f\sb {rest}(k)\sim [\tau\sb p\gamma(k)]^2$.
In the opposite case, when $1/\tau\sb p\gamma(k)$ is small,
$f\sb{com }(k)$ has corresponding smallness: $f\sb{com}(k)\sim
1/\tau\sb p\gamma(k)$. As a simple model of such a function we
adopt
\begin{equation} \label{eq:rest}
  f\sb {rest}(k)= 1- f\sb {com}(k) =[\tau\sb p\gamma(k)/ (1+
  \tau\sb p\gamma(k))]^2\ .
\end{equation}

Using \REF{rest}, we rewrite Eqs.~\Ref{eff-den} and \Ref{gamma-p1}
as Eqs.~\Ref{r-eff} and \Ref{gamma-p}.  Note that these equations,
which follow from the physical reasoning described above, give the
same expression for $\gamma\sb p(k)$ as \REF{gamma-p} in our
``derivation'' in Sec.~\ref{sss:k-friction}. We consider this fact
as a strong support of the physical relevance of our one-fluid
model for a turbulently flowing suspension given by
Eqs.~\Ref{NS-basic}-\Ref{nu-eff}, with $k$-dependent effective
density, fluid-particle damping frequency $\gamma\sb p$, and
effective kinematic viscosity~$\nu\sb{eff}(k)$.

%%%%%%%%%%%%%%%%%%%%%%%%%%%%%%%%%%%%%

%%%%%%%%%%%%%%%%%%%%%%%%%%%%%
\SS{ass}{Basic assumptions}

The theory developed in this paper is based on a number of
assumptions and simplifications described below:
\begin{enumerate}
\item All particles in the suspension are spheres with the same
density $\rho\sb p $ and the same radius $a$.

\item The radius of the particles is  small enough, see
Eq(\ref{eq:St-apll}).

\item The particle-particle interaction will be neglected,
assuming that the volume fraction $\psi\ll 1$. Nevertheless, for
the very heavy particles with $\rho\sb p \gg \rho \sb f $, the
mass loading $\phi$ may be of the order of unity, leading to a
significant modification of the turbulent flow  by particles.

\item The turbulent flow is stationary, homogeneous and isotropic.

\item In our equations for the energy balance~\Ref{budget} we will
use simple (but physically relevant) closure procedures based on
our effective (one-fluid) Navier-Stokes equation for
suspensions~\Ref{NS-basic} and on the Richardson-Kolmogorov
cascade picture of turbulence.

\end{enumerate}
%%%%%%%%%%%%%%%%%%%%%%%%%%%%%%%%%%%%%%%%%%%%%%%%
\SS{NSo-derivation}{Formal derivation of the effective NS equation
for suspensions}

In the derivation we begin with the NS \REF{NSf} for the fluid
component.  Instead of the averaged expression~\Ref{friction} for
the fluid-particle friction force we will use the following
detailed expression
%%%%%%%%%%%%%%
\begin{eqnarray}\label{eq:force}
\B f\sb p (t,\B r) =\sum_j \B F\sb p(t,\B r_j) \delta (\B r-\B r_j)\,,
\end{eqnarray}
%%%%%%%%%%%%%
in which $\B F\sb p (\B r_j,t)$ is the force between the fluid and
$j$-particle positioned at $\r=\r_j$. Assume [as in derivation of
Eqs.~\Ref{friction} and \Ref{NSp}] that the statistics of
particles is independent of the statistics of turbulence and,
moreover, that their distribution is space homogeneous. In that
case, we can replace the sum over the position of particles by a
space integration:
\[
\sum_j    \to \frac { 1} { \ell^3 }\int \d \B r_j \,,
\]
 where $\ell^3$ is the volume per particle. In this approximation
\begin{eqnarray}
\label{eq:Fp}
\B f\sb p  (\B r,t) = \B F\sb p  (\B r,t)  / \ell^3\ .
\end{eqnarray}
We compute $\B F\sb p \tr$ for small enough particles with a
radius $a$ satisfying inequality~\Ref{St-apll}, such that the
fluid flow in the vicinity of a particle may be considered as
laminar (assumption~\ref{ss:ass}-2). Then one can apply Stokes law
for the force $\B F\sb p \tr$:
\begin{eqnarray}\label{eq:Stokes-law}
  \B F\sb p \tr=\zeta\,[ \B v\sb p(t)- \B u\tr] \,
\end{eqnarray}
with the friction coefficient $\zeta$  for heavy particles (with
the density $\rho\sb p \gg \rho \sb f$)   is given by
\begin{eqnarray}
  \label{eq:fric}
  \zeta &=& 6\pi \,\rho \sb f\,\, \nu\, a \ .
\end{eqnarray}

The Newton equation for  a particle reads:
\begin{eqnarray} \label{eq:Newton}
m\sb p \frac{\d\, \B v\sb p (t)}{ \d t} = - \B F\sb p \tr =\zeta\,[
 \B u\tr  - \B v\sb p(t) ] \ .
\end{eqnarray}
A  formal solution of this equation
\begin{eqnarray}
  \label{eq:solution}
  \B v\sb p (t)=\Big[\tau\sb p \frac{\d }
{ \d t}+1 \Big]^{-1} \B u\tr \,,
\end{eqnarray}
allows one to express the force $\B F\sb p\tr$ as follows
\begin{eqnarray}
  \label{eq:Force}
  \B F\sb p \tr= m\sb p \frac{\d }{ \d t}\Big[\tau\sb p \frac{\d }{ \d
  t}+1 \Big]^{-1} \B u\tr\ .
\end{eqnarray}
Here $\tau\sb p$ is the particle response time:
\begin{eqnarray} \label{eq:t-Stokes}
\tau\sb p = m\sb p / 6 \pi \,\nu\, \rho \sb f \, a \ .
\end{eqnarray}
The total time derivative $(\d\,/\d t) $ as usual takes into
account the time dependence of the particle coordinate $\B r$:
\begin{eqnarray}
  \label{eq:total-1}
  \frac{\d }{ \d t}=\Big[ \pot{}+\B v\sb p (t)\cdot \B \nabla \Big]\ .
\end{eqnarray}
Due to the (immersed) particle inertia they do not follow
Lagrangian trajectories of fluid particles. Therefore, generally
speaking, $(\, \d\,/\d t\,) $ does not coincide with the total
Lagrangian time derivative in the fluid,
\begin{eqnarray}
  \label{eq:total}
  \frac{D}{D\, t}=\Big[ \pot{}+\B u\tr \cdot \B \nabla \Big]\ .
\end{eqnarray}
Consider the relationship between $(\d\,/\d t)$ and $(D/D\, t) $:
\begin{eqnarray}
\nn
 && \frac{D \B u\tr}{D\, t}= \frac{\d \B u\tr}{\d\, t}+[v\sb p^\alpha
 - u^\alpha \tr] \nabla_\alpha \B u\tr \\ \label{eq:D-d1}%%%%%%%%%%%%
&=&\frac{\d \B u\tr}{\d\,
 t}- \frac{\d }{\d t}\frac{\tau\sb p}{1+\tau\sb p \frac{\d }{\d
 t}} u\tr ^\alpha \nabla^\alpha_1 \B u_1 \br %%%%%%%%%%%%
&=&\frac{\d }{\d t}\frac{1}{1+\tau\sb p \frac{\d }{\d
 t}}\left[\Big(1+ \tau\sb p\frac{\p }{\p
 t}\Big)+ \hat {\C L}\right] \B u\tr \,,\\
&& \hskip -1cm
\hat {\C L} \, \B u \tr \equiv \tau\sb p\left[ (\B v_{\text p,1}
\cdot \B \nabla )\B u\tr - (\B u \cdot \B \nabla_1) \B u_1
\right]\ .\label{eq:L}
\end{eqnarray}
Here $\B u_1\equiv \B u(t_1,\B r_1)$, $\B \nabla_1=\d/ \d \B r_1$,
and all derivatives with respect $t_1$ and $\B r_1$ are taken at
$t_1=t$ and $\B r_1= \B r$. Together with \REF{total} this gives:
\begin{eqnarray}
  \label{eq:Force1}
  \B F\sb p \tr= \frac{D \B u\tr}{D\, t}\frac {1}{1+ \tau\sb p\frac{\p
  }{\p t} + \hat {\C L}}\, \B u\tr
\end{eqnarray}
For particles with a small response time Ferry and
Balachandar~\cite{01FB}  show, that the particle velocity depends
only on local fluid quantities (the velocity and its spatial and
temporal derivatives). They derive an expansion of the particle
velocity in terms of the particle response time which generalizes
those of previous researchers. For large values of the ratio of
the particle density and the fluid density and for small values of
the particle response time our Eq.~(\ref{eq:Force1}) for the force
on a particle gives the same equation for the particle velocity as
derived in Ref.~~\cite{01FB}.

 Substitution of Eq.~(\ref{eq:Force1}) into NS equation~\Ref{NSf} yields:
\begin{eqnarray}\nn
\rho\sb f \Big[ \pot {} &+& (\B u \cdot \B \nabla)\Big]
\Big\{1+\frac{\phi}{1+ \tau\sb p\big(\pot{} + \hat{\C L}
\big)}\Big\} \B u + \B \nabla p \\ \label{eq:NS1}
&=& \rho\sb f
\,\nu \nabla^2 \B u +\B f \,,
\end{eqnarray}
where $\phi$ is the mass loading parameter. For simplicity we
consider here only the case of heavy particles with negligibly
small volume loading parameter, $\psi\ll 1$. However, the mass
loading parameter may be of the order of unity. For example, for
the water droplets in the air, $\rho\sb{p}/\rho \sb{f}\approx
10^3$ and for $\phi=1$, the volume fraction $\psi\approx 10^{-3}$.

The inverse operator in Eq.~\Ref{NS1} may be understood as a
Taylor expansion with respect to the nonlinearity $(\B u \cdot \B
\nabla)$:
\begin{eqnarray}\nn
 \frac{\phi}{1+ \tau\sb p\big(\pot{} + \hat{\C L}
\big)}=  \frac{1}{1+\tau\sb p \pot{} } -   \frac{ \hat{\C L}  }
{\big(1+\tau\sb p \pot{}
\big)^2}+ \dots\\
\label{eq:Taylor}
  \end{eqnarray}
This expansion produces higher order [~in $(\B u \cdot \B
\nabla)$~] nonlinear terms in the effective NS equation~\Ref{NS1}.
These terms are not important for big eddies with $\tau\sb
p\gamma(k)\ll 1$ for which the operator in the braces in the LHS
of Eq.~\Ref{NS1}, $\{\dots \}$, is close to the factor $1+\phi$.
In the opposite case, for small scale eddies with $\tau\sb
p\gamma(k)\gg 1$ the operator $\{\dots\}=1$. Both limiting cases
one easily gets from the first term in the Taylor
expansion~\Ref{Taylor} in which there is no contribution from
$\hat{\C L}$.  It means that only for intermediate scales with
$\tau\sb p\gamma(k) \sim 1$ this operator may be quantitatively
important.  For a qualitative description of the ``transient''
process between these two regimes it is enough to account for the
first term of expansion~\Ref{Taylor}. In this approximation the
turbulent fluid velocity around the particle is approximated by
the velocity at a fixed coordinate, which is reasonable in
statistical sense and exact in the limit $\tau\sb p\gamma(k)\ll
1$. With this approximation Eq.~\Ref{NS1} turns into
\begin{eqnarray}\label{eq:NS2}
&& \hskip -1.5cm \hat \rho\sb {eff} \Big[ \pot {} + (\B u \cdot \B
\nabla)\Big] \B u + \B \nabla p = \rho\sb f \,\nu \nabla^2 \B u
+\B f \,,\\ \label{eq:rho-eff} && \hat \rho\sb {eff} \equiv
\rho\sb f \Big\{1+\frac{\phi}{\tau\sb p \pot{} +1}\Big\} \,,
\end{eqnarray}
where $ \hat \rho\sb {eff} $ may be considered as an operator of
effective density for a suspensions.

Since we are interested in the incompressible flows, we can project the
potential components out of the equation of motion. This may be done
by the projection operator $\BC P$, defined via its kernel
$\C P^{\alpha \beta} (\B r)$:
\begin{eqnarray}\label{eq:Pr}
\C P^{\alpha \beta}(\B r) &\equiv& \int \frac{\d^3 k}{( 2 \pi)^3}
P^{\alpha \beta} (\B k) \exp [ - i \B k \cdot \B r]\,,
\\ \label{eq:Pk}
P^{\alpha \beta}(\B k)&=& \delta_{\alpha \beta}
-k_\alpha k_\beta/k^2 \ .
\end{eqnarray}
The application of $\BC P$ to any given vector field $\B a (\B r)$
is nonlocal, and it has the form
\begin{eqnarray}
\label{eq:Pa}
[\B {\C P} \cdot \B a (\B r) ]_\alpha = \int d \B r_1 \C P^{\alpha
\beta} (\B r - \B r_1) a_\beta (\B r_1) \ .
\end{eqnarray}
Applying $\BC P$ to Eq.~\Ref{NS2} we find
\begin{eqnarray} \label{eq:NS3}
\hat \rho\sb {eff} \Big[ \pot {} + \BC P\cdot (\B u \cdot \B
\nabla)\Big] \B u = \rho\sb f \,\nu \nabla^2 \B u +\B f \,,
\end{eqnarray}
This equation together with the definition~(\ref{eq:rho-eff}) for
the operator of the effective density constitutes  a one-fluid
description of a turbulently flowing suspension. However, the
operator form of the effective density  is not convenient for
practical calculations. To overcome this inconvenience we will
derive below another form for the effective parameters of this
equation.

%%%%%%%%%%%%%%%%%%%%%%%%%%%%%%%%%%%%%%%%%%%%%%%%%%%%%%
\SS{NSk-derivation}{NS equation for suspensions with $k$-dependent
parameters}

In our analytical description  of space homogeneous, stationary
turbulence it is convenient to consider \REF{NS3} in the $(\B
k,\omega)$-representation:
\begin{eqnarray}
  \label{eq:NS-s}
\big \{\omega
\big[\tilde \rho\sb{eff}'(\omega)\big] - \i \rho\sb
f\big[\nu k^2+
 \tilde \gamma\sb p (\omega)\big] \big\} \tilde {\B u}\ok\br
  =-\BC N\{\B u, \B u\} _{\omega,\k}+\tilde {\B f}\ok\ .
\end{eqnarray}
Here
\begin{eqnarray} \label{eq:Fourier-o}
\tilde {\B u}\ok &=& \int \d t\,\d \r \, \B u \tr \exp (\i \,
\omega\, t+\i \k\cdot \r)\,,\\ \label{eq:den} %%%%%%%%%%%
\tilde
\rho\sb{eff}'(\omega)&=&\text{Re}\{\tilde{\rho}\sb{eff}(\omega)\}
=\rho\sb f \Big[1+\frac{\phi}{1+(\omega \tau\sb p)^2}\Big],
\\ \label{eq:gp-omega} %%%%%%%
\tilde \gamma\sb p
 (\omega)&=& \frac{\omega}{\rho\sb f} \,
 \text{Im}\big[\{\tilde{\rho}\sb{eff}(\omega)\big]\}=
 \frac{\phi\,\omega^2 \tau\sb p
 }{1+ (\omega \tau\sb p )^2}\,,\br
\rho\sb{eff}(\omega)&=&\rho\sb f \left[1+\frac{\phi}{1-\i\, \omega
\tau\sb p}\right]\,,
\end{eqnarray}
\begin{eqnarray}
\label{eq:NL} \BC N\{\B u, \B u\} _{\omega,\k}
 \equiv \left[ \tilde{\rho}\sb{eff}(\omega)\BC P
\cdot(\tilde {\B u} \cdot \B \nabla) \tilde {\B u}
\right]_{\omega,\k}\ .
\end{eqnarray}
 $ \BC N\{\B u, \B u\} _{\omega,\k}$ denote the nonlinear term
  in $\ok$-representation and frequency $\nu k^2$
describes the viscous damping.

The Navier-Stokes equation for suspensions~\Ref{NS-s} involves a
frequency-dependent effective density of suspensions,
$\tilde\rho'\sb {eff}(\omega)$ and $\omega$-dependent frequency $
\tilde \gamma\sb p (\omega)$ responsible for the damping due to
fluid-particle friction. To use standard closure procedures in the
statistical description of turbulence one needs frequency
independent coefficients in the basic equation of motion. On other
hand, these procedures may be applied to equations with
$k$-dependent coefficients. Therefore for further analysis it is
much more convenient to deal with a $k$-dependent effective
density $\rho\sb {eff} (k)$ of $k$-eddies. To relate these
functions we note that the $k$-eddies have a characteristic
frequency of motions, $\gamma(k)$ [related to their life-time
$\tau(k)$ by a simple relation $\gamma(k)\sim 1/\tau(k)$].

\SSS{k-dep}{$k$-dependent effective density of suspensions}

In the inertial interval of scales \REF{NS-s} must preserve the
total kinetic energy of a suspension $\C E$ if one neglects the
fluid-particle friction $\tilde \gamma\sb p (\omega)\to 0$. The
equation for $\C E$ may be written in terms of the density
$[\tilde{\rho}'\sb{eff}(\omega)$ and $\tilde{F}\ok$, the pair
correlation function of the $\ok$-Fourier components of velocity,
$\tilde u\ok$. Namely
\begin{eqnarray}\label{eq:Ek}
&& \C E=\Int{3}{k}\frac{\d \omega}{2\pi}\frac{\tilde {\rho}'\sb
{eff}(\omega)}{2} \tilde F\ok\,, \br &&\hskip -0.4cm (2\pi)^4
\delta(\B k-\B k_1) \delta(\omega-\omega_1) \tilde F\ok
\equiv\langle \tilde {\B u}\ok\cdot \tilde {\B u}\oke 1 \rangle \
.
 \end{eqnarray}
For isotropic turbulence $ \tilde F\ok=\tilde F(\omega,k)$ and
\REF{Ek} allows one to introduce the one-dimensional energy
spectrum of suspension, $\C E(k)$ according to:
\begin{eqnarray}\label{eq:E}
\C E&=& \int \frac{\d k}{2\pi}\,\C E(k)\,,\\  \label{eq:Eks1}
 \C E(k)&=& \Int{}{\omega}\frac{\tilde \rho'\sb
{eff}(\omega)}{2 \pi} k^2 \tilde F(\omega,k)\ .
\end{eqnarray}

Define a $k$-dependent effective density of suspension, which
gives the same one-dimensional spectrum $\C E(k)$ as the
$\omega$-dependent effective density $\tilde \rho'\sb
{eff}(\omega)$:
\begin{eqnarray}
  \label{eq:relation2}
 \rho\sb {eff}(k)=\frac{\int \tilde \rho\sb {eff}'(\omega) \tilde
 F(\omega,k) \, \d \omega }{\int \tilde F(\omega,k) \, \d \omega } \ .
\end{eqnarray}
Then, \REF{Eks1} takes the form
\begin{eqnarray}\label{eq:Eks2}
\C E(k)&=& \frac{\rho\sb {eff}(k)}{2 \pi} k^2  \Int{}{\omega}
\tilde F(\omega,k)\ .
\end{eqnarray}

Notice that
\begin{eqnarray}
  \label{eq:Fk}
   \Int{}{\omega}  \tilde F(\omega,k)=F(k)\,,
\end{eqnarray}
is the simultaneous velocity pair correlation function.
 With this notations
\REF{Eks2} may be written as
\begin{eqnarray}\label{eq:Eks3}
\C E(k)&=& \frac{\rho\sb {eff}(k)}{2 \pi} k^2F (k)\,,
\end{eqnarray}
while  the  traditional notation for 1-dimensional spectrum of
kinetic energy of fluid itself is $E(k)$:
\begin{eqnarray}\label{eq:Ek2}
E(k)&=& \frac{\rho\sb f}{2 \pi} k^2F(k)\ .
\end{eqnarray}
Formally speaking, in order to evaluate $ \rho\sb {eff}(k)$ by
\REF{relation2} we need to know the $\omega$ dependence of $F\ok$.
This is not a simple task. Instead we will use a few reasonable
forms of $F\ok$ and compare the resulting functions $ \rho\sb
{eff}(k)$. One of the frequently used is the Lorentzian form
\begin{eqnarray}
  \label{eq:omega-dep}
  \tilde F(\omega,k)= F (k)\frac{\gamma(k)/\pi
  }{\omega^2+\gamma^2(k)}\,,
\end{eqnarray}
which corresponds to the simplest ``one-pole'' approximation for the
Greens functions.  Using this $\omega$-dependence in
Eq.~\Ref{relation2} we have the following simple form for $\rho\sb
{eff}(k)$.
\begin{eqnarray}
  \label{eq:r-eff1}
   \rho\sb {eff}(k)&=&\rho\sb f \Big[1+\frac{\phi}{1+ \tau\sb p
   \gamma(k)} \Big]\ .
\end{eqnarray}
 For small $\tau\sb p \gamma(k)$ this gives a correction linear in
 $\tau\sb p \gamma(k)$
\begin{eqnarray}
  \label{eq:r-eff2}
   \rho\sb {eff}(k)\approx \rho\sb f \big[1- \phi \tau\sb p
   \gamma(k) \big]\,,
\end{eqnarray}
which contradicts the physical intuition. Indeed, one may consider
$k$-eddies as having  a motion with the characteristic frequency
$\gamma(k)$ and expect that $\rho\sb {eff}(k)$ may be obtained
from $\tilde \rho\sb {eff}'(\omega)$ with the substitution
$\omega\to \gamma(k)$. This gives a correction quadratic in
$\tau\sb p \gamma(k)$

\begin{eqnarray}
  \label{eq:r-eff3}
   \rho\sb {eff}(k)- \rho\sb f (1+\phi) \approx - \phi\, \rho\sb
   f\tau\sb p^2 \gamma^2(k) \ .
\end{eqnarray}
This contradiction follows from the fact  that the model function
\REF{omega-dep} decays very slowly for $\omega\to\infty$, like
$1/\omega^2$. It is known from the diagrammatic analysis of the
different time velocity correlation function $F(\tau,\k)$ that for
small $\tau$ the difference $F(\tau,\k)- F(0,\k)$ does not
contains $|\tau|$ and decays not slower than $\tau^2$.  Therefore
the Fourier transform of $F(\tau,\k)$, $\tilde F\ok$ has to decay
with $\omega$ faster than $1/\omega^2$, at least as $1/\omega^4$.
To meet this requirement we consider instead of \REF{omega-dep}
the function:
    \begin{eqnarray}
  \label{eq:omega-dep2}
  \tilde F(\omega,k)= F(k)\frac{2 \,\gamma^3(k)/\pi
  }{\big[\omega^2+\gamma^2(k)\big]^2}\,,
\end{eqnarray}
which gives instead of \REF{r-eff1}
\begin{eqnarray}
  \label{eq:r-eff4}
   \rho\sb {eff}(k)&=&\rho\sb f \Big\{1+\frac{\phi\, [1+ 2\, \tau\sb p
   \gamma(k)] }{\big[1+ \tau\sb p \gamma(k)\big]^2 } \Big\}\br
&=&\rho\sb f (1+\phi)  - \phi\,\rho\sb f \Big[\frac{\tau\sb p
   \gamma(k) }{1+ \tau\sb p \gamma(k)}\Big]^2 \ .
\end{eqnarray}
Now the correction to $ \rho\sb {eff}(k)$ is quadratic in $\tau\sb
p \gamma(k) $ which agrees with the expectation \Ref{r-eff3}.  One
observes the same agreement for any other model dependence $\tilde
F(\omega,k)$ decaying even faster than $1/\omega^4$.

Therefore the linear part of  \REF{NS-s} may be modelled as
\begin{eqnarray}
  \label{eq:NS-linear1}
 && \hskip -1.5cm \big \{\omega
 \rho\sb{eff}(k) - \i  \rho\sb f\big[ \nu k^2 + \tilde
 \gamma\sb p (\omega)\big] \big\} \tilde u\ok =\dots\,,
\end{eqnarray}
with $\rho\sb {eff}(k)$ given by \REF{r-eff4}.
%%%%%%%%%%%%%%%%%%%%%%%%%%%%%%%%%%%%%%%%%%%%%%
\SSS{k-friction}{Effective fluid-particle damping frequency $\gamma\sb
  p(k)$}

Using \REF{NS-s} or \REF{NS-linear1} together with \REF{Eks1} we
can compute the contribution of the fluid-particle friction to the
damping of $\C E(k)$:
\begin{eqnarray}
  \label{eq:p-dam1}
  \pot{\C E(k)}\Big |\sb p = - 2 \rho\sb f \Int{}{\omega}\frac{\tilde
  \gamma \sb p(\omega)}{2 \pi} k^2 \tilde F(\omega,k)\ .
\end{eqnarray}
Introduce an $\omega$-independent fluid-particle damping friction by
a standard relation:
\begin{eqnarray}
  \label{eq:p-dam2}
  \pot{\C E(k)}\Big |\sb p = - 2  \gamma\sb p (k)\C E(k)\ .
\end{eqnarray}
Combining these two equations with \REF{Eks1} one gets
\begin{eqnarray}
  \label{eq:relation3}
 \gamma \sb p(k)=\frac{\rho\sb f\int \tilde \gamma \sb p (\omega)
 \tilde F(\omega,k) \, \d \omega }{\int \tilde \rho\sb {eff}'(\omega)
 \tilde F(\omega,k) \, \d \omega } \ .
\end{eqnarray}
Substitution of Eqs.~\Ref{gp-omega} and \Ref{omega-dep} into
\REF{relation3} gives \REF{gamma-p} for $ \gamma\sb p(k)$.
With this knowledge, \Ref{NS-linear1} may be further simplified as
follows
\begin{eqnarray}
  \label{eq:NS-linear2}
 && \rho\sb{eff}(k) \big \{\omega
 - \i\big[\nu\sb{eff}(k)\, k^2 +
 \gamma\sb p (k)\big] \big\} \tilde {\B u}\ok \br
 &&=- \BC N\{\B u, \B u\} _{\omega,\k} +\tilde{\B f}\ .
\end{eqnarray}
Here $\nu\sb{eff}(k)$ is given by \REF{nu-eff}.  Notice that this
equation gives the same dissipation rate \Ref{p-dam2} due to
fluid-particle friction as \REF{NS-s} and the same dissipation
rate
\begin{eqnarray}
  \label{eq:nu-dam2}
  \pot{\C E(k)}\Big |_\nu  = - 2 \nu\sb{eff}k^2  \C E(k)
\end{eqnarray}
due to the kinematic viscosity.

The suggested form of $\BC N\{\B u, \B u\} _{\omega,\k}$ in terms
of ${\rho}\sb{eff}(k)$ will be discussed in the following Sec.
~\ref{sss:NLk}.

%%%%%%%%%%%%%%%%%%%%%%%%%%%%%%%%%%%%%%%%%%%%%%%%%%%%%%%%%%%%
\SSS{NLk}{$\omega$-independent nonlinearity of the
  effective NS equation}
\paragraph{Nonlinearity  in the usual NS equation.}

Consider first the nonlinear term in the ``usual'' NS equation for
single-phase flow. In $\ok$-representation it has the form (see
\eg ~Refs.~\onlinecite{B8,B4}):
\begin{eqnarray}
  \label{eq:NL1}
&& \C N\{\B u, \B u\}_{\omega,\k}^\alpha
 = \int\frac{\d^3 k_1\, \d^3 k_2 \,\d \omega_1\, \d
  \omega_2} {(2\pi)^4 } \br
&\times&  \gamma^{\alpha\beta\gamma}_{\B k\,\B k_1\B
  k_2} u_\beta^*\oke 1 \tilde u_\gamma^* \oke 2
  \delta(\omega+\omega_1+\omega_2) \ .
\end{eqnarray}
Here $\gamma^{\alpha\beta\gamma}_{\B k\,\B k_1\B k_2}$ is the so
called {\em vertex of interaction} given by~\REF{vertex}. It
includes transversal projectors accounting for the
incompressibility of the fluid, delta function of $\B k$-vectors
originating from the space homogeneity of the problem and is
proportional to $k$ (as a reflection of $\B \nabla$ operator in
the nonlinear term in $\B r$-representation).

The vertex $ \gamma^{\alpha\beta\gamma}_{\B k\,\B k_1\B k_2} $
satisfies so-called \it{Jacobi identity}
\begin{eqnarray}
\label{eq:Jac}
\gamma^{\alpha\beta\gamma}_{\B k\,\B k_1\B k_2} +
\gamma^{\gamma\alpha\beta}_{\B k_2\B k\,\B k_1}+
\gamma^{\beta\gamma\alpha}_{\B k_1\B k_2\B k}=0\,,
\end{eqnarray}
as a consequence of the energy conservation by the Euler equation.

\paragraph{Nonlinearity in the effective NS \REF{NS-basic}.}
A rigorous derivation of the nonlinear term in the effective NS
\REF{NS-basic} is a very delicate issue. For example, in \REF{NS2}
we used the operator of the effective density~\Ref{rho-eff} containing
only the first term of expansion~\Ref{Taylor}. This approximation does
not account for all terms, second orderin $\B u $. This derivation is
beyond the scope of this paper. Instead we present here physical
arguments which allows us to propose a form of $\BC N\{\B u, \B
u\}_{\omega,\k}$ which satisfies all needed requirements.

By analogy with \REF{NL1} we can write:
\begin{eqnarray}
  \label{eq:NL2}
  &&\C N\{\B u, \B u\}_{\omega,\k}^\alpha= \int\frac{\d^3 k_1\, \d^3
  k_2 \,\d \omega_1\, \d \omega_2} {(2\pi)^4 } \br &\times&
  \Gamma^{\alpha\beta\gamma}_{\B k\,\B k_1\B k_2} u_\beta^*\oke 1
  \tilde u_\gamma^* \oke 2 \delta(\omega+\omega_1+\omega_2) \,,
\end{eqnarray}
where the vertex $\Gamma^{\alpha\beta\gamma}_{\B k\,\B k_1\B
  k_2}$ differs from $\gamma^{\alpha\beta\gamma}_{\B k\,\B k_1\B
  k_2}$, \REF{vertex}, because now $\rho\sb f\ne \rho\sb{eff}(k)$.

The simplest possible generalization of the vertex, just a
replacement $\rho\sb f\to \rho\sb {eff}(k)$ in \REF{vertex} leads
to a vertex $\Gamma^{\alpha\beta\gamma}_{\B k\,\B k_1\B k_2}$,
which does not satisfy the Jacobi identity
\begin{eqnarray}
\label{eq:Jac1}
\Gamma^{\alpha\beta\gamma}_{\B k\,\B k_1\B k_2} +
\Gamma^{\gamma\alpha\beta}_{\B k_2\B k\,\B k_1}+
\Gamma^{\beta\gamma\alpha}_{\B k_1\B k_2\B k}=0\,,
\end{eqnarray}
leading to violation of the  conservation of the kinetic energy of
suspension $\C E$. We suggest \REF{vertex-basic} for
$\Gamma^{\alpha\beta\gamma}_{\B k\,\B k_1\B k_2}$.  Clearly, due
to \REF{Jac} this vertex satisfies requirement~\Ref{Jac1} and
consequently, \REF{NS-basic} conserves the energy $\C E$.

Another physical requirement is Galilean invariance of the
problem. This is the case for the standard NS equation with
vertex~\Ref{vertex} in which $\rho\sb f$ is $k$-independent.  For
the  $k$-dependent density in the vertex~\Ref{vertex-basic}
Galilean invariance implies that   in the limit, when one of the
wave-numbers is much smaller then two others (say $k_1\ll
k_2\equiv k_3$), the effective density must depend on the smallest
$k$-vector. Obviously, this is the case for the $k$-argument of
$\rho\sb {eff}(k)$ in \REF{vertex-basic}.  This guarantees
Galilean invariance of \REF{NS-basic}.

Now  \REF{NS-linear2} involves only  $\omega$-independent
coefficients and may be rewritten  in $\tk$ representation,
see~\REF{NS-basic}.

%%%%%%%%%%%%%%%%%%%%%%%%%%%%%%%%%%%%%%%%%%%%%%%%%%%%%%%
\S{balance}{Budget of kinetic energy in turbulent suspensions}

In this section we consider the budget of kinetic energy in
turbulent suspension. In Sec.~\ref{ss:energy} we will use the
one-fluid model~\Ref{NS-basic} to derive (for homogeneous,
isotropic turbulence) the following \it{budget} equation for the
(1-dimensional) density of kinetic energy:
\begin{eqnarray}\label{eq:budget}
&& \frac{\p \C E(t,k)}{2\,\p t}+\big[\gamma_0(k) +\gamma\sb
 p(k)\big] \C E(t,k) \br  &=&\C W (t,k)+\C J(t,k)\ .
\end{eqnarray}
The left hand side (LHS) of this equation includes two damping
terms, $\gamma_0(k)\C E(t, k)$, caused by the effective kinematic
viscosity  and $\gamma\sb p(k)\C E(t,k)$ caused by the
fluid-particle friction. The density $\C E(t,k)$ is given by
\REF{CEk}. The right hand side (RHS) of \REF{budget} includes the
source of energy $\C W(t,k)$, localized in the energy containing
interval, and the energy redistribution term, $\C J (t,k)$.

The budget equation~\Ref{budget} is exact, but unfortunately not
closed.  Equations \REF{nu-eff} for the effective kinematic viscosity
and \REF{gamma-p} for $\gamma\sb p(k)$ includes ``turnover frequency''
of $k$-eddies, $\gamma(k)$. Also $\C W(t,k)$ contains yet unknown
$(u,f)$ correlations, \REF{W}. And finally $\C J(t,k)$ is given by
Eqs.~\Ref{CJk}, \Ref{bal2} via 3-rd order velocity correlations
$F_3$. There are many reasonable closure procedures for the
approximation of high-order velocity correlations by lower-order ones.
To elucidate the basic physics of the problem at hand, in this paper
we will use the simplest possible closure. Application of more
sophisticated closures will be done elsewhere.

%%%%%%%%%%%%%%%%%%%%%%%%%%%%%%%%%%%%%%
\SS{energy}{The energy budget equation}

In order to derive \REF{budget} we multiply \REF{NS-basic} by $\B
u (t,\B k')$ and average:
\begin{eqnarray}\nn
      &&\rho\sb{eff}(k)\Big\{\frac{\p F (t,\k)}{2\,\p t}
 +\big[\gamma_0(k)+\gamma \sb p(k)\big]F (t,\k)\Big\} \\ \label{eq:bal1}
&& =J\tk + W\tk\,,\\
    \label{eq:bal2}
 &&\hskip -1.15cm  J\tk\equiv \int\frac{\d^3 k_1\, \d^3 k_2 }
 {(2\pi)^3} \Gamma^{\alpha\beta\gamma}_{\B k\,\B k_1\B
  k_2}  F_3^{\alpha\beta\gamma}(t;\k,\k_1,\k_2)\ .
\end{eqnarray}
Here $F \tk$, and $F_3(t;\dots)$ are the 2nd and 3rd order
\it{simultaneous} velocity correlation functions taken at
\it{overall} time $t$:
 \begin{eqnarray}
\label{eq:F2} &&(2\pi)^3 \delta(\B k+\B k_1) F (t;\B k)= \langle
\B u\tk\cdot \B u
(t,\k_1) \rangle \,, \\
\label{eq:F3}
&&(2\pi)^3 \delta(\B k+\B k_1+\B k_2)
F_3^{\alpha\beta\gamma}(t;\B k,\B k_1,\B k_2)\br
&&= \langle  u^\alpha\tk  u^\beta (t,\k_1) u^\gamma (t,\k_2)
\rangle \ .
\end{eqnarray}
Note that the time $t$ in the argument of $F(\k)$ in
Eqs.~(\ref{eq:Fk}) -- (\ref{eq:omega-dep}) is omitted: $F \tk =
F(\k)$.

In \REF{bal1} we introduce also the \it{simultaneous}
$(u,f)$-cross correlation functions
 \begin{eqnarray}\label{eq:W}
 (2\pi)^3 \delta(\B k-\B k_1) W \tk=\<\B u\tk \B f(t,\k_1) \>\ .
\end{eqnarray}
We can rewrite \REF{Eks2} for the  density of the kinetic energy
of suspension in terms of $F \tk=F (t,k)$ (for isotropic
turbulence):
\begin{eqnarray}
\label{eq:CEk} \C E(t,k)=\frac{\rho\sb {eff}(k)}{2 \pi} k^2 F
(t,k)\ .
\end{eqnarray}
Multiplying Eq.~(\ref{eq:bal2}) by $ k^2 /2\pi$ one gets finally the
balance \REF{budget} in which
\begin{eqnarray}
  \label{eq:CWk}
  \C W(t,k)=\frac{ k^2 }{2 \pi} W(t,k)\,, \\
 \label{eq:CJk}
  \C J(t,k)=\frac{ k^2 }{2 \pi}  J(t,k)\ .
\end{eqnarray}
Notice that effective vertex $ \Gamma^{\alpha\beta\gamma}_{\B
k\,\B k_1\B k_2}$ in \REF{bal2} was constructed such that the
total kinetic energy $\C E$ is the integral of motion (neglecting
pumping and damping):  $\int_0^\infty \C J(t,k) \d k=0$. Therefore
the energy redistribution term  $\C J(t,k)$ may be written in the
divergent form:
\begin{eqnarray}
  \label{eq:flux}
  \C J(t,k)&=&- \frac{\d \ve(t,k)}{\d k}\,,\qquad \text{where}\\
\label{eq:flux1}
\varepsilon (t,k)&=& \int_k^\infty \d k\, J(t,k)
\end{eqnarray}
is the (one-dimensional) energy flux over scales.

In the rest of the paper we will consider only stationary solutions of
\REF{budget}. Omitting (here and below) time argument one finally has:
\begin{eqnarray}\label{eq:s-budget}
\big[\gamma_0(k) +\gamma \sb  p(k)\big] \C E(k)=\C W (k) - \frac{\d
\ve(k)}{\d k}\ .
\end{eqnarray}

%%%%%%%%%%%%%%%%%%%%%%%%%%%%%%%%%%%%%%
\SS{closure}{Simple closure for the energy budget equation}

The effective density in our approach, \REF{r-eff}, depends on the
characteristic  frequency $k$-eddies $\gamma(k)$. This object may
be evaluated as the inverse life  time of these eddies which is
determined by their viscous damping and energy loss in the cascade
processes. Accordingly, $\gamma(k)$ is a sum of two terms
\begin{equation}\label{eq:sum-g}
\gamma(k)=\gamma_0(k)+\gamma\sb c (k)\,,
\end{equation}
where $\gamma_0(k)$, \REF{nu-eff}, is the viscous frequency and
$\gamma\sb c (k)$ may be evaluated as the turnover frequency of
$k$-eddies:
\begin{eqnarray}
\label{eq:turn-time}
  \gamma\sb c (k)\sim k \, U_k\,, \quad \text{where} \quad U_k\sim \sqrt{{k
  \, \C E(k)}\big/{\rho\sb {eff}(k)}}
\end{eqnarray}
is the characteristic velocity of  $k$-eddies. We therefore define
\begin{eqnarray}
  \label{eq:gamma-k}
   \gamma\sb c(k)=C_\gamma \sqrt{{k^3  \, \C E(k)}\big/{\rho\sb {eff}(k)}} \,,
\end{eqnarray}
where $C_\gamma$ is some  dimensional constant, presumably of the
order of unity.  Clearly, the same evaluation~\Ref{gamma-k} one
gets from a dimensional reconstruction of $\gamma(k)$ in terms of
the only relevant (in the K41 picture of turbulence)  variables
$k$, $\C E(k)$ and $\rho\sb {eff}(k)$.

In the same manner, by the dimensional reasoning, one gets the
following evaluation of the energy flux:
\begin{eqnarray}
  \label{eq:e-flux}
  \varepsilon (k)= C_\ve \sqrt{{k^5 \, \C E^3(k)}\big/{\rho\sb
  {eff}(k)}}\,,
\end{eqnarray}
where $ C_\ve =O(1)$.  Notice that in pure fluids [with $\rho\sb
{eff}(k)\to \rho\sb f$] Eqs.~\Ref{gamma-k} and \Ref{e-flux} are
nothing but the K41 evaluation of the corresponding objects. This
become even more transparent  if one rewrites Eq.~\Ref{e-flux} in
the  more familiar form:
\begin{eqnarray}
  \label{eq:energy3}
   \C E(k) =C_1 \big[\varepsilon ^2 (k)\, \rho\sb{eff}(k)\big]^{1/3}
   k^{-5/3}\,, \  C_1\equiv C_\ve^{-2/3}.
\end{eqnarray}
Together with~\REF{gamma-k} this gives a  useful evaluation of
$\gamma\sb c(k)$ via $\ve(k)$:
\begin{eqnarray}
  \label{eq:gamma2}
  \gamma\sb c(k)=C_2\Big[\frac{\ve(k)\,k^2 }{\rho\sb{eff}(k)}\Big]^{1/3}\,,\
C_2\equiv \frac{C_\gamma}{C_\ve^{1/3}}\ .
\end{eqnarray}

Lastly, we have to evaluate the energy input in the system $\C
W(k)$. It follows from~\REF{NS-basic} that $u(k)$ may be evaluated
as $f(k)/\gamma(k) \rho\sb{eff}(k)$. Together with Eqs.~\Ref{W},
\Ref{CWk} and \Ref{gamma-k} this gives:
\begin{eqnarray}
  \label{eq:CW}
  \C W(k)= C_w f^2_k\sqrt{\frac{k\, \C E(k)}{\rho\sb{eff}(k)}}\,,
\end{eqnarray}
where $ C_w =O(1)$ and $ f^2_k$ is the pair
correlation of the forcing [which may be defined similarly
to~\Ref{W}].

%%%%%%%%%%%%%%%%%%%%%%%%%%%%%%%%%%5
\SS{basic-eq}{Dimensionless budget equation}

For the convenience of the reader we present here the full set of
equations which will be studied below.  To non-dimensionalize this
equation we define   a dimensionless wave-number, $\kappa$, and the
integral-scale related parameters
\begin{equation}\label{eq:int-s} \kappa =k\,L\,, \
\ve_{_L}=\ve\Big(\frac{1}{L}\Big)\,, \
\gamma_{_L}=\gamma\Big(\frac{1}{L}\Big)\,, \ \rho_{_L}=
\rho\sb{eff}\Big(\frac{1}{L}\Big)\,,
\end{equation}
Define also the dimensionless functions
\begin{eqnarray}\label{eq:def-non}
\ve_\kappa&= &\frac{\ve(k)}{\ve_{_L}}\,,\qquad \gamma_\kappa=
\frac{\gamma(k)}{\gamma_{_L}}\,,\br %%%%%%%%%%%
\rho_\kappa&=& \frac{\rho\sb{eff}(k)}{\rho_{_L}} \,,
\quad \C W_\kappa=
\frac{\C W(k)}{\C W(1/L)}\,,
\end{eqnarray}
in which the argument $\kappa$ is written as a subscript to
distinguish them from the corresponding dimensional functions of
the dimensional argument $k$.

 The resulting dimensionless budget equation reads
\begin{eqnarray}\label{eq:bud2}
\frac{\d \ve_\kappa}{\d y}&+&\frac{\ve_\kappa}{y}C\, T_\kappa \br
&+&\frac{C_1}{\Re\sb
s}\left(\frac{y\,\ve_\kappa^2}{\rho_\kappa^2}\right)^{1/3}(1+T_\kappa)
=\C W_\kappa\,,\\ \label{eq:Ty} %%%%%%%%%%%%%%%%%%%%%%%%%%%%
T_\kappa&\equiv &\frac{\phi\, \delta\,
\gamma_\kappa}{(1+\phi)(1+2\,\delta\, \gamma_\kappa) +(\delta\,
\gamma_\kappa)^2}\,, \br %%%%%%%%%%%%%%%%%%
C&=&C_1\,C_2\,, \qquad \delta=\gamma_{_L}\tau\sb p\ .
\end{eqnarray}
Here we used \REF{energy3} and defined the  Reynolds numbers for
the carrier fluid, $\Re\sb f$, and the effective Reynolds number for
the suspension $\Re\sb s$
\begin{eqnarray}\label{eq:Re}
\Re\sb f &=&\frac{L\,v_{_L}}{\nu} \,, \qquad v_{_L}=\left(\frac{
\ve_{_L}\, L }{\rho _{_L}}\right)^{1/3}\,, \br %%%%%%%%%%
Re\sb s &=&\frac{L\,v_{_L}}{\nu_{_L}} \,, \qquad
 \nu_{_L} \equiv
\nu\sb{eff}(L^{-1})= \nu \, \frac{\rho_{_L}}{\rho \sb f}\,,\\
\label{eq:ratio1} %%%%%%%%%%%%%%%%%%
\rho_{_L}&=& \rho \sb f\left[1+\phi \frac{1+2\,
\delta}{(1+\delta)^2}\right]\ .
\end{eqnarray}
 in terms of  the rms turbulent velocity, $v_{_L}$, dominated by
$L$-eddies. Obviously, $\Re\sb f$ involves the  kinematic viscosity of
the  carrier fluid, $\nu$, while $\Re\sb s$ depends on the
effective kinematic viscosity of the suspension,
$\nu\sb{eff}(L^{-1})$,  for the outer scale of turbulence,  $L$.

\REF{bud2} has to be considered together with equations for
$\rho_\kappa$ and $\gamma_\kappa$, which follows from
Eqs.~\Ref{r-eff}, \Ref{sum-g} and \Ref{gamma2}:
\begin{eqnarray}\nn
\rho_\kappa&=&\left[ 1+\phi \frac{1+2\, \delta \gamma_\kappa }
 {(1+\delta\,\gamma_\kappa)^2} \right]\Bigg /
 \left[ 1+\phi \frac{1+2\, \delta}
 {(1+\delta)^2} \right]\,, \\ \label{eq:rel-1}
 \gamma_\kappa&=& \frac{\kappa^2}{C_2\, \Re\sb s\,\rho_\kappa}
 +\frac{\ve_\kappa^{1/3}\,
 \kappa^{2/3}}{\rho_\kappa^{1/3}}\ .
\end{eqnarray}
These two equations allow us to express the functions
$\gamma_\kappa$ and $\rho_\kappa$ in terms of $\ve_\kappa$. With
these solutions, \REF{bud2} becomes an ordinary differential
equation for the only function $\ve_\kappa$.

 The first line of \REF{bud2} describes the effect of particles in the
 inertial  integral of scales. This part involves  the mass loading, $\phi$,
 the dimensionless particle response time (normalized by the eddy life time),
  $\delta$, and the parameter $C$, characterizing our version of the K41 closure.

The second line of \REF{bud2} represents the effect of the viscous
friction, which is proportional to $1/\Re\sb s$, and the pumping
term $\C W_\kappa$, which we choose as follows:
\begin{equation}\label{eq:Wy}
\C W_\kappa=\frac{1}{\sqrt{2\pi}\sigma}\exp \left[-\frac{(y-1)^2}{2\,
\sigma^2}\right]\ .
\end{equation}
This function has a maximum at $y=1$ (the input of  energy is
largest at $k=1/L$), while the parameter $\sigma$ describes the
characteristic width of the pumping region. In addition, the
function $\C W_\kappa$ satisfies the normalization constrain
\begin{equation}\label{eq:norm}
\int\limits_{-\infty}^{\infty}\C W_\kappa\,\d y=1\,,
\end{equation}
which follows from  \REF{bud2} in the limit $\sigma \ll 1$.
%%%%%%%%%%%%%%%%%%%%%%%%%

\section{Solution of the budget equation}
\label{s:prelim}
%%%%%%%%%%%%%%%%%%%%%%%%%%%%%%%%%%%%%%%%
\subsection{Simplification of the energy pumping term}
\label{ss:pump}
 First notice, that the turbulence statistics in the energy
containing range, $k\, L= y\sim 1$ is not universal and depends on
the type of energy pumping, in our case, on the function $\C
W_\kappa$. Therefore for general analysis, which is not aimed at
the study of some particular type of turbulence generation, we can
take the pumping of energy in a narrow shell in the $\k$-space.
This means
\begin{equation}\label{eq:lim}
\lim_{\sigma\to \, 0}\left \{ \C W_\kappa \right \}=\delta
(\kappa)\ ,
\end{equation}
where $\delta(\kappa)$ is the Dirac $\delta$ function. In this
limit
 and with zero boundary conditions for $\ve_\kappa$,
$\gamma_\kappa$ at $\kappa=0$ (and, consequently, $\rho_\kappa=1$
at $\kappa=0$) , \REF{bud2} can be solved on the interval $0\le
\kappa \le 1$. This gives
\begin{eqnarray}\label{eq:bound}
\ve_\kappa=1\,,\qquad \gamma_\kappa=1\,, \qquad  \rho_\kappa=1\,,
\quad \text{at}\   \kappa =1\ .
\end{eqnarray}
In the limit~\Ref{lim}, \REF{bud2} has zero RHS for $\kappa >1$:
\begin{equation}\label{eq:bud3}
\frac{\d \ve_\kappa}{\d \kappa }+\frac{\ve_\kappa}{\kappa}C\,
T_\kappa +\frac{C_1}{\Re\sb s}\left(\frac{\kappa
\,\ve_\kappa^2}{\rho_\kappa^2}\right)^{1/3}(1+T_\kappa) =0\,,
\end{equation}
Relations \Ref{bound} can be considered as the boundary conditions
for \REF{bud3} at $\kappa=1$.

\subsection{Particle free case and limit of small particles}
\label{ss:no-part}

Consider now the particle-free case, $\phi=0$, and the case of
very small particles, $\delta=0$,  for finite $\Re\sb s $. In both
cases \REF{bud3} becomes:
\begin{eqnarray}\label{eq:bud4}
\frac{\d \ve_\kappa}{\d y}+\frac{C_1\,\kappa^{1/3}
\ve^{2/3}}{\Re\sb s }=0\ .
\end{eqnarray}
We took here in account that according to \REF{rel-1}
$\rho_\kappa=1$ for  $\phi=0$ and also for $\delta=0$. Notice,
that for $\phi=0$, $\nu_{_L}=\nu$ and, consequently, $\Re\sb
s=\Re\sb f$, while for $\delta=0$, $\nu_{_L}=\nu/(1+\phi)$ and
$\Re\sb s=\Re\sb f(1+\phi)$. The reason is that for $\delta\to 0$
all particles are fully involved in turbulent motions and one can
consider a suspension as a single but heavier fluid  with the
density $\rho\sb f(1+\phi)$.

The solution of \REF{bud4} with the boundary condition $\ve_1=1$
is
\begin{eqnarray}\label{eq:sol1}
\ve_\kappa =\Big[1+\frac{C_1}{4\Re\sb
s}\left(1-\kappa^{4/3}\right)\Big]^3 \ .
\end{eqnarray}
In the particle free case $\phi=0$ and   this solution turns into
\begin{eqnarray}\label{eq:sol0}
\ve_\kappa ^{(0)}=\Big[1+\frac{C_1}{4\Re\sb
f}\left(1-\kappa^{4/3}\right)\Big]^3 \,,
\end{eqnarray}
where $\Re\sb s\Rightarrow \Re\sb f$ as we discussed above.  In
the bulk of the inertial interval these solutions give a small
viscous correction to the K41 solution with the constant energy
flux $\ve_\kappa=1$. Namely
\begin{equation} \label{eq:sol2}
\ve_\kappa \approx 1+3\,C_1 \,(1- \kappa^{4/3})\big / (4\Re\sb
s)\,, \quad \text{for} \quad \kappa \ll 1/\Re\sb s\ .
\end{equation}
The local in the $k$-space closure procedure,  used in the paper,
works reasonably well in the inertial interval, where the energy
exchange between eddies is dominated by the eddies of compatible
scales. However, it is violated in the bulk of the viscous
subrange, where the dynamics of eddies of very small scales is
dominated not by their self-interaction, but by  their energy
exchange with $\eta$-eddies of the Kolmogorov micro-scale $\eta$.
Therefore we cannot expect \REF{bud3} to reproduce the exponential
decay of the energy flux in the viscous subrange. Nevertheless,
this equation describes on a qualitative level the behavior of the
energy flux until the very end of the inertial interval giving the
crossover scale to the viscous subrange, i.e. the value of $\eta$.
According to \REF{sol1}, the energy flux become zero at
\begin{equation}\label{eq:kappa-cr}
    \kappa\sb{cr}= (1+ 4\,\Re\sb s/C_1)^{3/4}\ .
\end{equation}
It convenient to introduce here an effective Reynolds number of
the carrier fluid and suspensions
\begin{equation}\label{eq:Re-eff}
\Re\sp{eff}\sb f=4\Re\sb f/C_1\,,\quad \Re\sp{eff}\sb s=4\Re\sb s
/C_1\,,
\end{equation}
which enters in the corresponding Kolmogorov-41 evaluations of the
viscous cutoff. For example for the fluid
\begin{equation}\label{eq:eta}
  \eta\sb f=L /  \kappa\sb{cr}\approx L \big /[\Re \sp{eff}\sb f]^{3/4}\ .
\end{equation}
%%%%%%%%%%%%%%%%%%%%%%%%%%%%%%%%%
\subsection{Iterative solution in the inertial interval}
\label{ss:energy-bud}

In the bulk of the inertial interval, after neglecting the viscous
terms (i.e. for $\Re\sb s\to \infty)$,   \REF{bud3} becomes
\begin{eqnarray}\label{eq:bud}
\frac{\d \ve_\kappa}{\d
\kappa}&+&\frac{\ve_\kappa}{\kappa}\frac{C\, \phi\, \delta\,
\gamma_\kappa}{(1+\phi)(1+2\,\delta\, \gamma_\kappa) +(\delta\,
\gamma_\kappa)^2}=0  \,, \\ \label{eq:rel-2} %%%%%%%%%%
\gamma_\kappa^3 \,\rho_\kappa&=& \ve_\kappa \kappa^2\,,\br %%%%%%%%%%%%
 \rho_\kappa&=&\left[ 1+\phi
\frac{1+2\, \delta \gamma_\kappa }
 {(1+\delta\,\gamma_\kappa)^2} \right]\Bigg /
 \left[ 1+\phi \frac{1+2\, \delta}
 {(1+\delta)^2} \right] \ .
\end{eqnarray}

%%%%%%%%%%%%%%%%%%%%%%%%%%%%%%%%%%%%%
\subsubsection{Large scale  solution of the budget
equation}

%%%%%%%%%%%%%%%%%%%
In region of large scales,  $\kappa\approx 1$, the functions
$\rho_\kappa\approx 1$  and we can simplify \REF{rel-2} by the
replacement  $\rho_\kappa \Rightarrow 1$ in the equation for
$\gamma_\kappa$, i.e. $\gamma_\kappa \Rightarrow \ve_\kappa^{1/3}
\kappa^{2/3}$. In the denominator of \REF{bud}, where the
$\kappa$-dependence of $\gamma_\kappa$ is less essential,  we can
make further simplification, replacing  $\gamma_\kappa \Rightarrow
\kappa^{2/3}$. The resulting equation allows separation of
variables:
\begin{eqnarray}
  \label{eq:bud1}
 \frac{2}{\delta} \frac{\ \, \d\,  \ve_\kappa ^{-1/3}}{ \d\,
\kappa^{2/3}}&=& C\, \Psi _0(\kappa)\,, \br %%%%%%%%%%%%%
\Psi_0(\kappa)&=& \frac{\phi  } {(1+\phi)(1+ 2 \delta \kappa^{2/3}
) + \delta ^2 \kappa^{4/3}}\ .
\end{eqnarray}
The solution of this equation with the boundary conditions $\ve_1=1$
is $\ve_\kappa=\ve_{0,\kappa}$, where
\begin{eqnarray}\label{eq:ve0}
\ve_{0,\kappa}&=& \frac{ 1}{\big[1+C\, J_0(\kappa)\big]^3}\,,\br %%
J_0(\kappa)&=& \frac{\delta }{2}\int \limits _1^\kappa \Psi
_0(x)\,
\d\, x^{2/3}  \br %%%%%%%%%%%%%
&=& \frac{\sqrt{\phi}}{4\sqrt{1+\phi}}
 \Bigg\{\ln\left[\frac{\delta \kappa^{2/3}+1+\phi - \sqrt{\phi(1+\phi)} }
{\delta +1+\phi - \sqrt{\phi(1+\phi)} } \right] \br %%%%%%%%%%%%%
&& \hskip 1.4cm - \ln\left[\frac{\delta \kappa^{2/3}+1+\phi +
\sqrt{\phi(1+\phi)} } {\delta +1+\phi + \sqrt{\phi(1+\phi)} }\right] \Bigg\} \,.  %%%%%%
\end{eqnarray}
Now with \REF{rel-2} we find the following approximations
\begin{eqnarray} \nn %%%%%%%%%%
\rho_{0,\kappa}&=&\left[ 1+\phi \frac{1+2\, \delta\,
\ve_{0,\kappa}^{1/3}\, \kappa^{2/3} }
 {\big(1+\delta\,\ve_{0,\kappa}^{1/3}\, \kappa^{2/3}\big)^2} \right]\Bigg /
 \left[ 1+\phi \frac{1+2\, \delta}
 {\big(1+\delta\big )^2} \right] \,,\\ \label{eq:gamma0}  %%%%%%%%%%%%%%%%
 \gamma_{0,\kappa} &=& \big( \ve_{0,\kappa} \kappa^2/\rho
 _{0,\kappa}\big)^{1/3}\ .
\end{eqnarray}

Formally speaking, the analytical solution
(\ref{eq:ve0}--\ref{eq:gamma0}) is valid only for $\kappa \approx
1$. To find the solution of the initial \REF{bud}  in the whole
interval of scales, we will iterate this equation, taking the
analytical solution as a starting function. Comparing in
Sec.~\ref{ss:accuracy} this solution with the next order
iterations and with the numerical solution of \REF{bud}, we will
find an actual region of applicability of the analytical solution
(\ref{eq:ve0}--\ref{eq:gamma0}).

\subsubsection{First improvement and subsequent iterations}
\label{sss:iter}

With the  analytical solution (\ref{eq:ve0}--\ref{eq:gamma0}), we can
improve approximation \Ref{bud1} of \REF{bud} by
$\rho_\kappa\Rightarrow\rho_{0,\kappa}$ [instead of
$\rho_\kappa\Rightarrow~1$], which gives $\gamma_\kappa
\Rightarrow \ve_\kappa ^{1/3}\kappa^{2/3} /\rho_{0,\ve}$ in the
numerator of \REF{bud}. In the denominator we replace
$\gamma_\kappa \Rightarrow \ve_{0,\kappa}^{1/3}\kappa^{2/3}
/\rho_{0,\ve}$. The improved simplification of \REF{bud} reads:
\begin{eqnarray}
  \label{eq:bud5}
\frac{2}{\delta} \frac{\ \, \d\,  \ve_\kappa ^{-1/3}}{ \d\,
\kappa^{2/3}}&=& C\, \Psi _1(\kappa)\,, \br %%%%%%%%%%%%
\Psi _1(\kappa)&=&  \frac{\phi}
{\rho_{0,\kappa}^{1/3}\big[\big(1+\phi)(1+ 2 \delta \,
\gamma_{0,\kappa}\big) + \big(\delta\, \gamma_{0,\kappa}\big) ^2
\big]}\ .
\end{eqnarray}
Integration of this equation gives the first iterative solution of
\REF{bud}, $\ve_{\kappa}=\ve_{1,\kappa}$, where
\begin{eqnarray}\label{eq:ve1}
\ve_{1,\kappa}&=& \frac{ 1}{\big[1+C\, J_1(\kappa)\big]^3}\,,\br %%
J_1(\kappa)&=& \frac{\delta }{2}\int \limits _1^\kappa \Psi
_1(x)\,
\d\, x^{2/3}  \ .  %%%%%%
\end{eqnarray}
This allows further improvement of approximations \Ref{gamma0}
\begin{eqnarray} \nn %%%%%%%%%%
\rho_{1,\kappa}&=&\left[ 1+\phi \frac{1+2\, \delta\,
\gamma_{0,\kappa} }
 {\big(1+\delta\,\gamma_{0,\kappa}\big)^2} \right]\Bigg /
 \left[ 1+\phi \frac{1+2\, \delta}
 {\big(1+\delta\big)^2} \right] \,,\\ \label{eq:gamma1}  %%%%%%%%%%%%%%%%
 \gamma_{1,\kappa} &=& \big( \ve_{1,\kappa} \kappa^2/\rho
 _{1,\kappa}\big)^{1/3}\ .
\end{eqnarray}
Now, the  next iteration steps are obvious. The $n$-order solution
is
\begin{eqnarray}\label{eq:ven}
\ve_{n,\kappa}&=& \frac{ 1}{\big[1+C\, J_n(\kappa)\big]^3}\,,\br %%
J_n(\kappa)&=& \frac{\delta }{2}\int \limits _1^\kappa \Psi
_n(x)\,
\d\, x^{2/3} \,, \br  %%%%%%
\Psi _n(\kappa)&=&  \frac{\phi}
{\rho_{n-1,\kappa}^{1/3}\big[\big(1+\phi)(1+ 2 \delta \,
\gamma_{n-1,\kappa}\big) + \big(\delta\, \gamma_{n-1,\kappa}\big)
^2\big]
}\,, \br %%%%%%%%%%
\rho_{n,\kappa}&=&\left[ 1+\phi \frac{1+2\, \delta\,
\gamma_{n-1,\kappa} }
 {\big(1+\delta\,\gamma_{n-1,\kappa}\big)^2} \right]\Bigg /
 \left[ 1+\phi \frac{1+2\, \delta}
 {\big(1+\delta\big)^2} \right] \,,\\ \label{eq:gamman}  %%%%%%%%%%%%%%%%
 \gamma_{n,\kappa} &=& \big( \ve_{n,\kappa} \kappa^2/\rho
 _{n,\kappa}\big)^{1/3}\ .
\end{eqnarray}

%%%%%%%%%%%%%%%%%%%%%%%%%%%%%%%%
 \begin{figure}
\epsfxsize=7cm \epsfbox{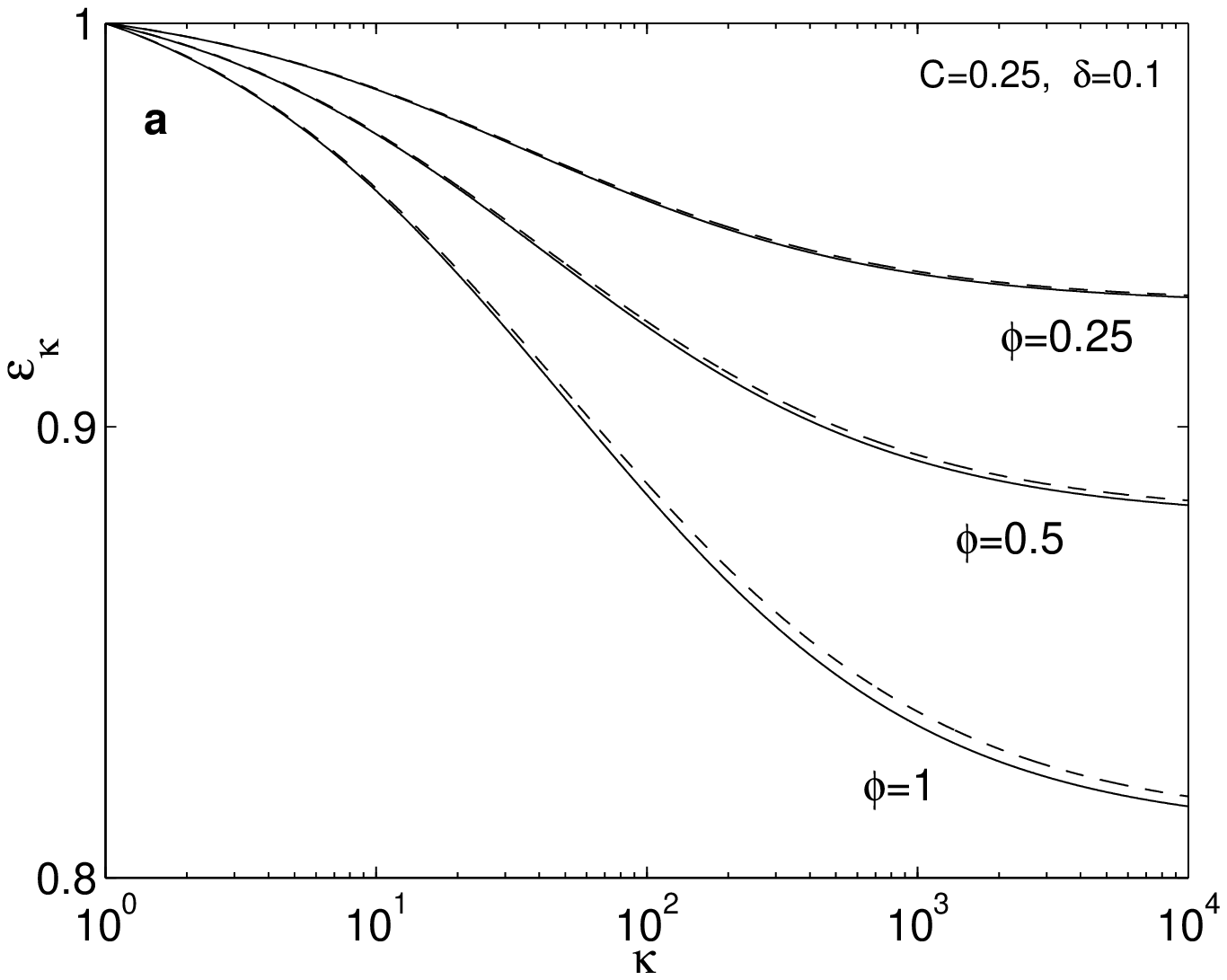} \epsfxsize=7cm
\epsfbox{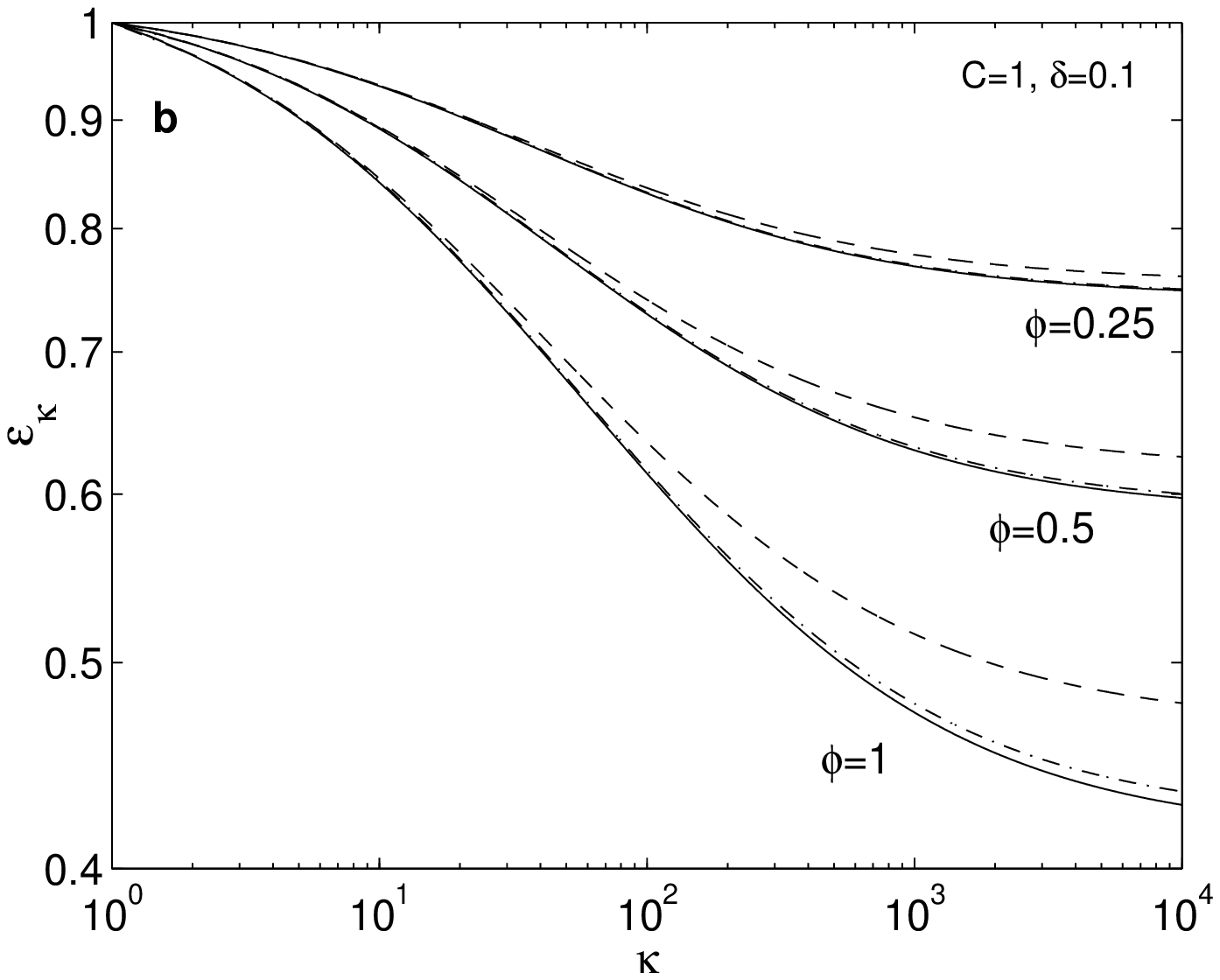} \epsfxsize=7cm
 \caption{\label{f:F1}  Log-Log plots of analytical solution,
$ \ve_{0,\kappa}$, (dashed lines), first iterative solution $
\ve_{1,\kappa}$, (dot-dashed lines) and "exact" numerical solution
(solid lines) for $\delta=0.1$ and various values $\phi$.  Panel
$a$: $C=0.25$, panel $b$: $C=1$.}
\end{figure}
%%%

\subsection{Accuracy of the iterative solutions}
\label{ss:accuracy}

To get an understanding of the accuracy of the analytical solution
$\ve_{0,\kappa}$,  \REF{ve0}, and the first iterative solution
$\ve_{1,\kappa}$, we compare them with the  "numerically exact"
solutions of \REF{bud}, $\ve_\kappa$ in the wide inertial range of
four decades.

We found  that for all values of $\kappa$ and $\delta$ the analytical
function $\ve_{0,\kappa}$ works unexpectedly well for $C\,\phi \le
0.25$. To illustrate this, we plot in Fig.~\ref{f:F1} functions
$\ve_{0,\kappa}$, $\ve_{1,\kappa}$  and $\ve_{\kappa}$ for
$C=0.25$ (panel $a$) and $C=1$ (panel $b$)  for $\phi=0.25,\, 0.5$
and $\phi=1$ with $\delta=0.1$. The relative difference between
$\ve_{0,\kappa}$ and $\ve_{\kappa}$ is about a few percents for
all three cases in panel $a$ and for the case $\phi=0.25$ in panel
$b$.

For larger values of the product $C\,\phi$ the accuracy of  few
percents is achieved in the smaller region of $\kappa$, where $\tau\sb
p\gamma(k)=\delta \kappa^{2/3}<1$, i.e. approximately for $\kappa
\leq \delta^{-3/2}$. For example for $\delta=0.01$ this is three
decades, $\kappa<10^3$, while  for $\delta=0.1$ only for
$\kappa<30$, as we show in Fig.~\ref{f:F1}$b$. Moreover, the first
iterative solution, $\ve_{1,\kappa}$ gives a very good
approximation to $\ve_{\kappa}$ for all reasonable values  of
parameters. This is illustrated in Fig.~\ref{f:F1}$b$ for $C=1$
and $\phi=1$. Notice, that for $C=0.25$ and $\phi=1$
(Fig.~\ref{f:F1}$a$) the plots of $\ve_{1,\kappa}$ $\ve_{\kappa}$
are undistinguished within the line width.

The conclusion is that for the qualitative and semi-quantitative
description of the turbulence modification by particles in the
inertial interval we can use the analytical solution
(\ref{eq:ve0}--\ref{eq:gamma0}), corrected, if needed, by the
first iteration, $\ve_{1,\kappa}$.

%%%%%%%%%%%%%%%%%%%%%%%%%%%%%%%%%%%%%%%%%%%%%%%%%%%%%%%%%
\section{Turbulence modification by particles}
\label{s:modification}
\subsection{Preliminaries}
\label{ss:prelim}

Consider now separately the density of kinetic energy of the
carrier fluid, $E\sb f(k)$,  and that of the particle, $E\sb p
(k)$ (i.e. the density of the kinetic energy of the particle
velocity field).  According to Eqs. \Ref{r-eff}, \Ref{Eks3},
\Ref{Ek2} and \Ref{energy3}
\begin{eqnarray}\label{eq:Ef}
E\sb f(k) &=&C_1 \rho\sb f \, \big[ \varepsilon (k)\big /
\rho\sb{eff}(k)\big]^{2/3}k^{-5/3}\,, \\ \label{eq:Ep}
 E\sb p (k)&=& \phi\,\frac{1+2\, \tau\sb p \,
 \gamma(k)}{[1+ \tau\sb p\, \gamma(k)]^2}\, E\sb f (k)\ .
\end{eqnarray}
It is convenient to introduce the dimensionless functions of
$\kappa=k\,L$, $ E\sp f _\kappa$  and $ E\sp p _\kappa$, both
normalized by $E\sb f(L^{-1})$:
\begin{eqnarray}
\label{eq:nond} E\sp f _\kappa=\frac{E\sb f(k) }{E\sb
f(L^{-1})}\,,\quad E\sp p _\kappa=\frac{E\sb p(k) }{E\sb
f(L^{-1})}\,,
\end{eqnarray}
which may be written as
\begin{eqnarray}\label{eq:Efn}
E\sp f _\kappa &=&\big( \varepsilon_\kappa \big / \rho_\kappa\big)
 ^{2/3}\kappa ^{-5/3}\,, \\ \label{eq:Epn}
 E\sp p _\kappa &=& \phi\,\frac{1+2\, \delta \,
 \gamma_\kappa}{[1+ \delta \, \gamma_\kappa ]^2}\,E\sp f _\kappa \ .
\end{eqnarray}
Next, introduce the dimensionless ratio
\begin{eqnarray}
  \label{eq:ratio}
  R(\kappa) &\equiv& \frac{E\sp f _\kappa }{E^{0,\rm f} _\kappa }=
  \left[\frac{\ve_\kappa }{ \ve_\kappa ^{(0)} \rho_\kappa}
  \right]^{2/3}\,,
\end{eqnarray}
where $ E^{0,\rm f} _\kappa=[\ve_\kappa
^{(0)}]^{2/3}\kappa^{-5/3}$ is the density of turbulent kinetic
energy and $\ve_\kappa ^{(0)}$ is the energy flux in the
particle-free case, \REF{sol0}. The ratio $R(\kappa)$ is larger
(smaller) than unity in the case of enhancement (suppression) of
the turbulent energy by particles.

\subsection{Energy flux}
\label{ss:en-flux}

Our model  with local  in $k$ space parametrization of the energy
flux involves the parameter of the closure procedure $C$, which has to
be considered as a fit parameter which may be evaluated, for
example, by comparison with the direct numerical simulation. Generally
speaking, it is expected to be of the order of unity. For
simplification of the qualitative analysis of the effect of
particles on the statistics of turbulence we choose usually
$C=1/4$, for which we can use the analytical solution
(\ref{eq:ve0}--\ref{eq:gamma0}). The effect of $C$ on the behavior of
the function $\ve_\kappa$ is qualitativly described by the
approximation $\ve_\kappa (C)\sim 1 + C\,[\ve_\kappa(C-1)]$. This  is
illustrated by comparison of plots $\ve_{0,\kappa}$ for $C=0.25$ and
$C=1$  in Figs.~\ref{f:F1}$a$ and \ref{f:F1}$b$.

The plots in  Fig.~\ref{f:F1} also demonstrate the expected fact
that for small $\phi$ the effect of the particles is proportional to
$\phi$. This may  also clearly be seen from the balance \REF{bud}.
The balance equation itself also reflects a less trivial fact of
saturation of the effect of the particles in the limit $\phi\gg 1$;
the beginning of this saturation is clearly seen in
Fig.~\ref{f:F1}. The  physical reason for the saturation is that the
main  governing parameter in the problem is the ratio of the
particle energy to the energy of the \emph{suspension} but not to
the energy of the \emph{carrier fluid}. As an example of the
quantitative description of the effect of saturation consider
\REF{ve0} for $\ve_{0,\kappa}$ in the limit $\kappa\to
\infty$:
\begin{equation}\label{eq:ve-inf}
    \ve_{0,\infty}=\left[1+\frac{C\sqrt{\phi}}{4\sqrt{1+\phi}}
    \ln\left(
\frac{\delta+1+\phi +\sqrt{\phi(1+\phi)}}
{\delta+1+\phi-\sqrt{\phi(1+\phi)}}
  \right)
 \right]^{-3}.
\end{equation}
Evidently:
\begin{eqnarray}\label{eq:ve-limits}
 \ve_{0,\infty}&\approx &\left[1+\frac{C\phi}{2(1+\delta)}
 \right]^{-3}\,,\quad\text{for}\ \phi\ll 1\,,\br
\ve_{0,\infty}&\approx &\left[1+\frac{C}{4}\ln
\left(\frac{2\phi}{\delta+1}\right)
 \right]^{-3}\,,\quad\text{for}\ \phi\gg 1\ .
\end{eqnarray}

\begin{figure}
\epsfxsize=8cm \epsfbox{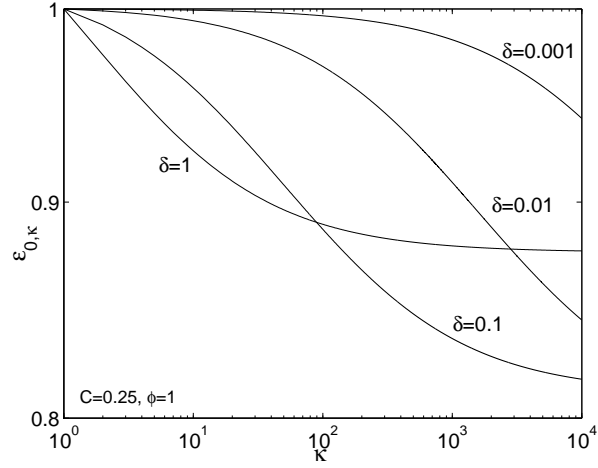}
\caption{\label{f:F2} Log-Log plots of analytical solution
$\ve_{0,\kappa}$ for  $\delta=1, $ $\ 0.1, \, 10^{-2}, $ $\,
10^{-3}$ (with $C=0.25$ and  $\phi=1$). }
\end{figure}

One sees in Fig.~\ref{f:F1} that the increase in the mass loading
$\phi$, leads to the suppression of the energy flux for large
$\kappa$ ( small scales). The onset of this suppression shifts to
smaller $\kappa$ (larger scales) with increasing $\delta$,
Fig.~\ref{f:F2}. To understand this, we note that $\tau \sb p
\gamma_\kappa \approx \delta \, \kappa^{2/3}$ is an important
governing parameter in the energy budget equation. Consequently,
with increasing particle response time, the fluid-particle
friction dissipate energy in the larger scale region. As is
evident from this figure, the main dissipation of energy occurs in
the region $0.1<\delta\kappa^{2/3} < 10$. Therefore,  for
$\delta>0.1 $ the total loss in the energy decreases with further
increase in $\delta$.

%%%%%%%%%%%%%%%%%

  %%%%%%%%%%%%%
  \subsection{Suppression and enhancement of turbulence in the inertial interval}
  \label{ss:inertial}

As we discussed in Sec.~\ref{ss:prelim} the effect of particles on
the energy distribution in suspension (with respect to the
particle free case) may be characterized by the ratio $R(\kappa)$,
\REF{ratio}. This effect is twofold: the fluid-particle friction
leads to suppression of the energy flux with increasing $\kappa$.
Accordingly, $\ve_\kappa$  in the numerator of Eq. \Ref{ratio}
decreases towards larger $\kappa$. On the other hand, for larger
$k$ less particles are involved in the motion and the effective
density, $\rho_\kappa$, decreases with $\kappa$. The factor
$\ve_\kappa^{(0)}=1$ in the limit $\Re\sb s\to \infty$. Therefore
in the inertial interval of scales  the behavior of $R(\kappa)$ is
defined by the strongest $\kappa$-dependence either of $\ve_\kappa$
or of $\rho_\kappa$. As we discussed, for $C<0.25$ it is sufficient to
use  the analytical solutions for  $\ve_{0,\kappa}$, \REF{ve0}, and
$\gamma_{0,\kappa}$, \REF{gamma0}. This gives:
\begin{eqnarray}\nn
 R_0(\kappa)=\left\{\frac{\ve_{0,\kappa}\big[1
+\phi\big(1+2\delta\big)\big/\big(1+\delta\big)^2\big]}
{1+\phi\big(1+2\delta\gamma_{0,\kappa}\big)\big/
\big(1+\delta\gamma_{0,\kappa}\big)^2 }\right\}^{2/3}\ .\\
\label{eq:Rkappa}
\end{eqnarray}
%%%%%%%%%%%%%%%%%%%%%%%%%%%%%
\begin{figure}
\epsfxsize=7.5cm \epsfbox{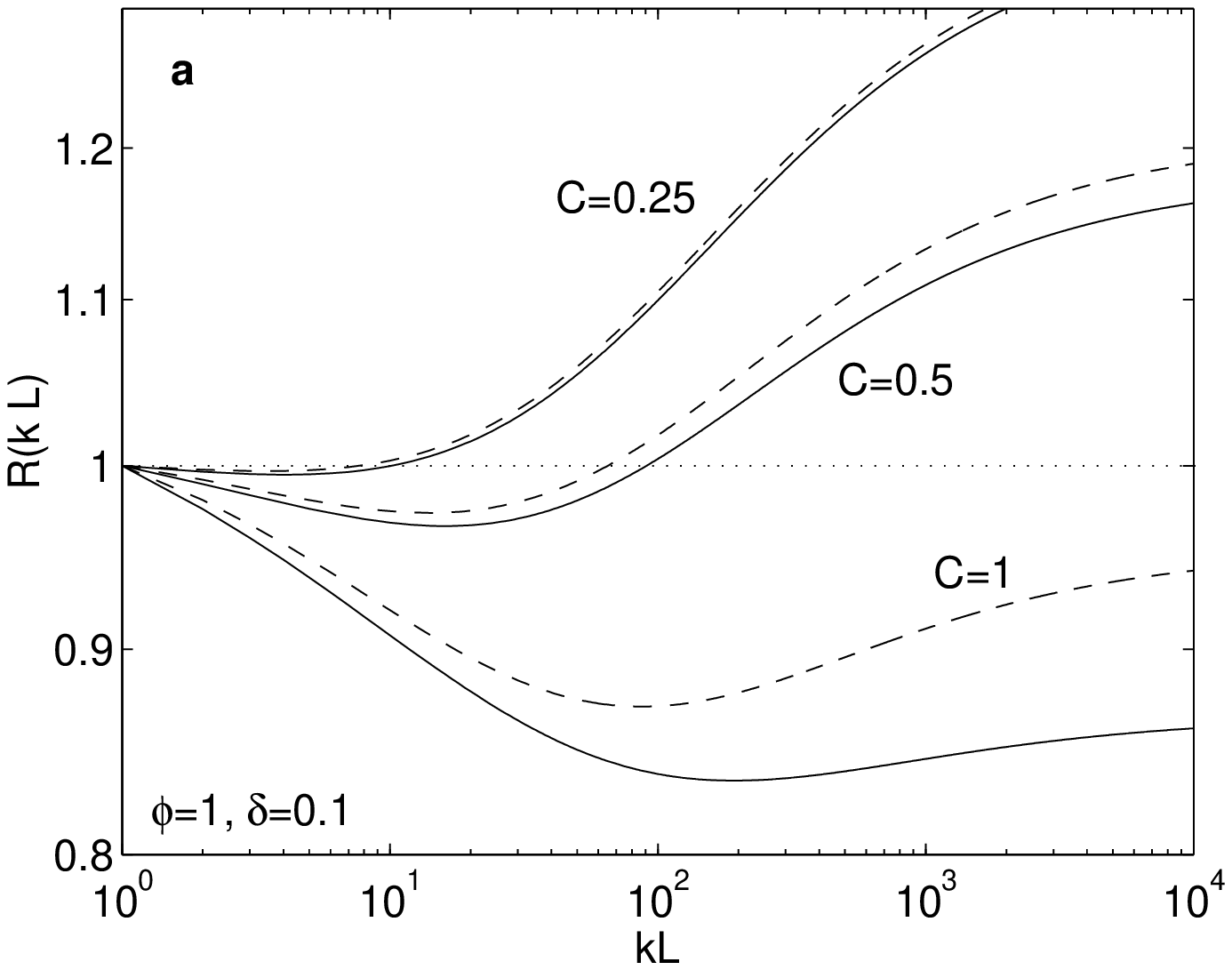} \epsfxsize=7.5cm
\epsfbox{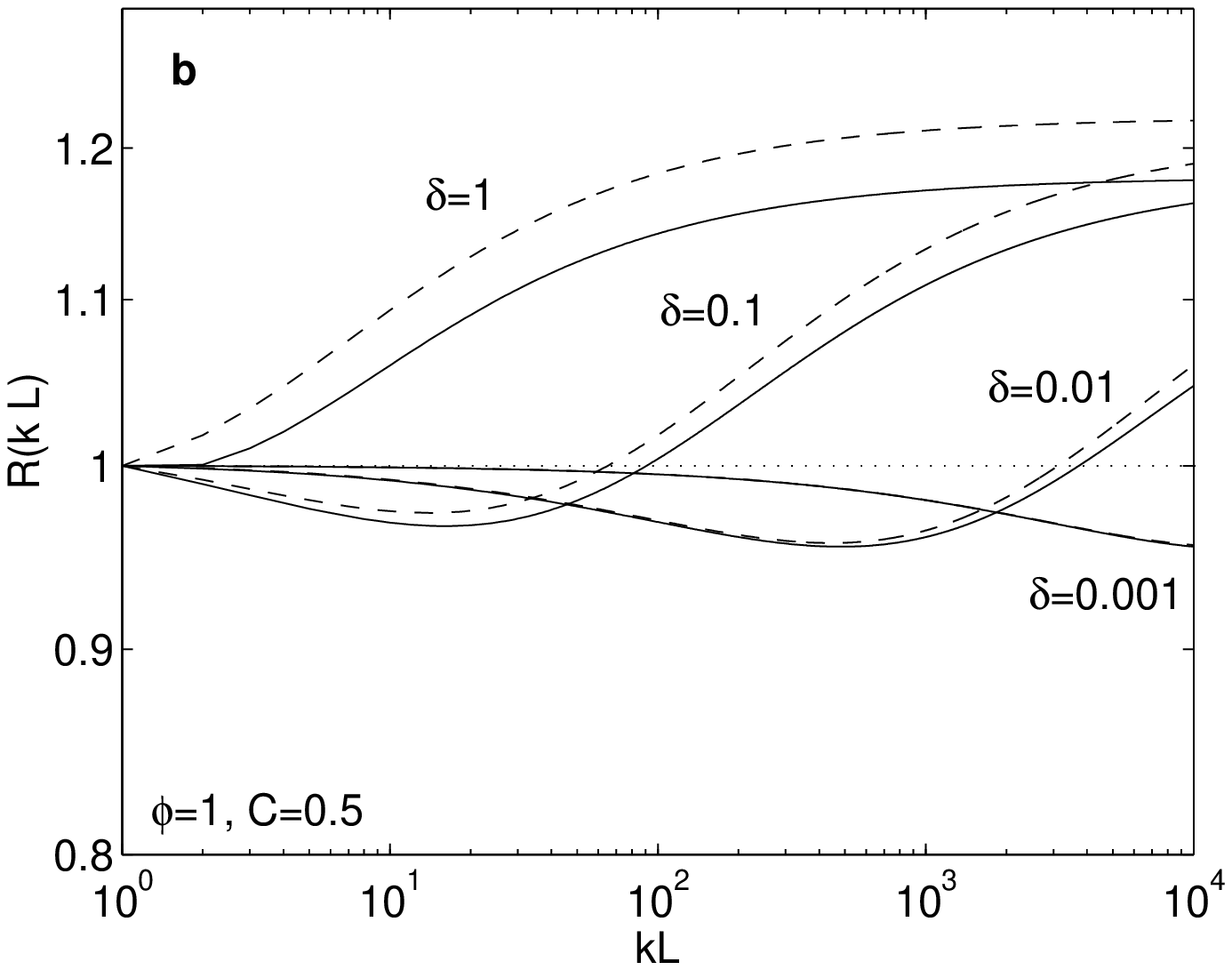} \caption{\label{f:F3} Log-log plots of the
analytical prediction $R_0(\kappa)$, \REF{Rkappa}, (dashed lines)
and the numerical result, $R(\kappa)$, for $\phi=1$ (remind, $\kappa=kL$). Panel $a$:
$\delta=0.1$ and  values of $C$ indicating corresponding lines.
Panel $b$:  $C=0.5$ and values of $\delta$   corresponding
lines. }
\end{figure}
%%%%%%%%%%%%%%%%%%%%%%%%%%%%%%%%%%%

Fig.~\ref{f:F3}$a$  demonstrates  how the  ratio $R(\kappa)$ depends on
the fit parameter of our model, $C$, which appeared in the budget
\REF{bud} in the front of the term, responsible for the
fluid-particle friction. Clearly, the relative importance of the
fluid-particle friction (with respect to the effect of the density
variation) increases with the value of $C$. In particular, for $C=1$
the friction dominates and $R(\kappa)<1$, for $C=0.25$ the density
variation dominates and $R(\kappa)>1$. For $0.25<C<1$, the density of
energy of the carrier fluid is suppressed for smaller $\kappa$ and
enhanced towards larger $\kappa$.  As is clearly seen in the figure, the
function $R(\kappa)$ has a minimum around some critical value $\kappa
\sb {cr}$ which depends on $C$ and $\kappa$. For $C\approx 1$ the
value of $\kappa \sb {cr}$ agrees with the expected estimate $\tau\sb
p
\gamma(k\sb {cr})\equiv \delta\,  \gamma_{\kappa\sb {cr}}\approx
1$.  With decreasing $\delta$ the position of the crossover ( and
of the minimum) is shifted towards larger $\kappa$. It is evident
that for $\kappa < \kappa \sb {cr}$ the effect of the
fluid-particle friction wins, while for $\kappa > \kappa\sb {cr}$
the effect of the decrease in the effective density of the
suspension is stronger. For $\kappa \gg \kappa\sb {cr}$ the
function $R(\kappa)$ saturates.

Notice that for $C<0.5$ the analytical prediction (dashed lines) is
pretty  close to the "exact" numerical result (solid lines)
indicating the qualitative validity of the analytical description of
the effect of particles on the energy distribution in suspensions
(within the model limitations). Therefore, one can find the
limiting value of $R_\infty \equiv R(\kappa\to\infty)$ from
\REF{Rkappa}:
\begin{eqnarray}
  \label{eq:R2inf}
  && R_{0,\infty}=\left[1+\frac{\phi\,(1+2\,\delta)}{(1+\delta)^2}
  \right]^{2/3}\br %%%%%%%%%%%%
  &&\times
  \left[1+\frac{C}{2}\sqrt{\frac{\phi}{1+\phi}}\ln\left(
\frac{\delta+1+\phi +\sqrt{\phi(1+\phi)}}
{\delta+1+\phi-\sqrt{\phi(1+\phi)}}
  \right)
 \right]^{-2}\ .
\end{eqnarray}
The analysis of this equation shows that the largest possible
enhancement of the turbulent energy in the inertial interval is
achieved for $\delta \approx 1$ and increases with $\phi$.

%%%%%%%%%%%%%%%%%%%%%%%%%%%%%
\begin{figure}
\epsfxsize=7.5cm \epsfbox{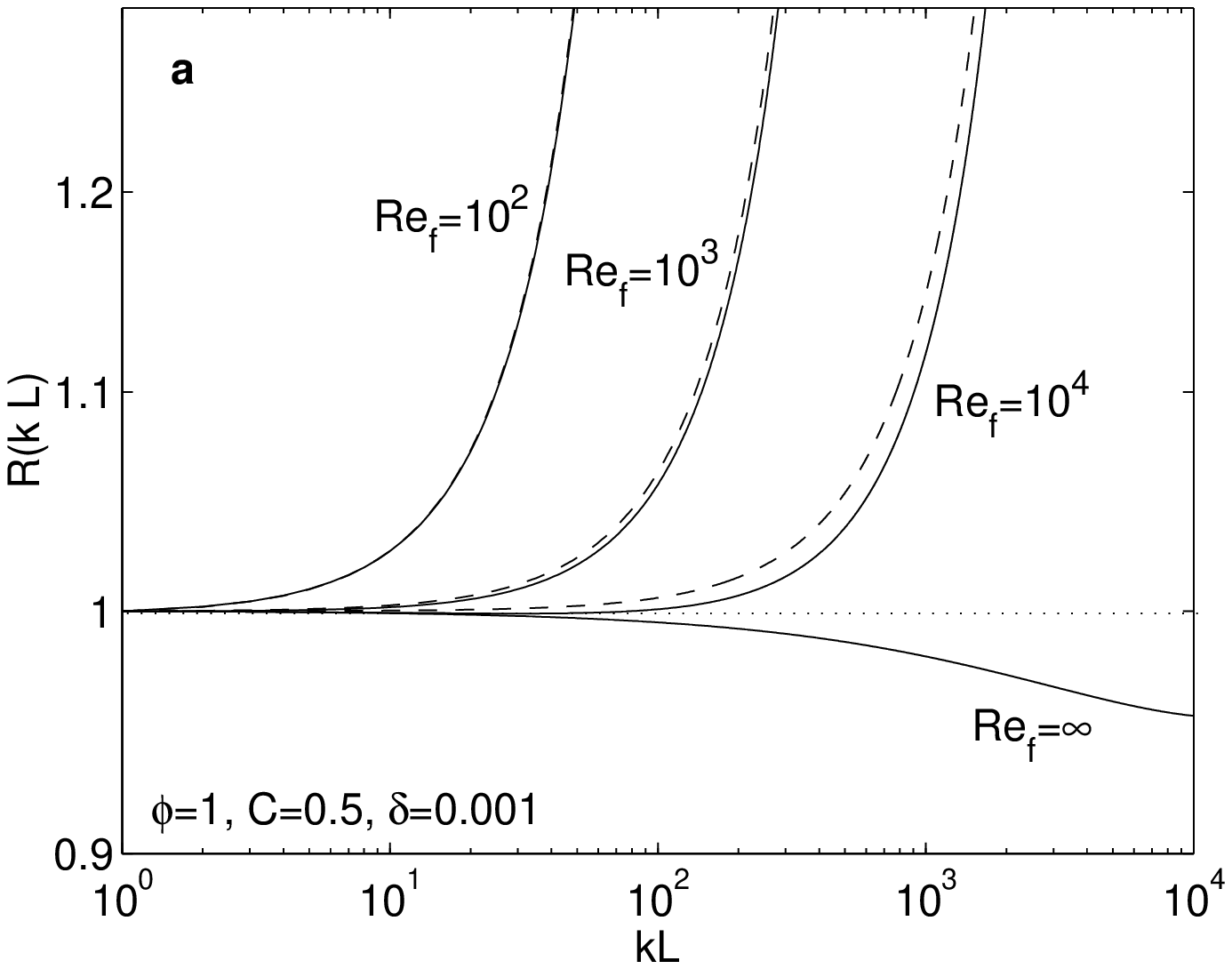}\\%%%%%%%%%%%%%%%
\epsfxsize=7.5cm \epsfbox{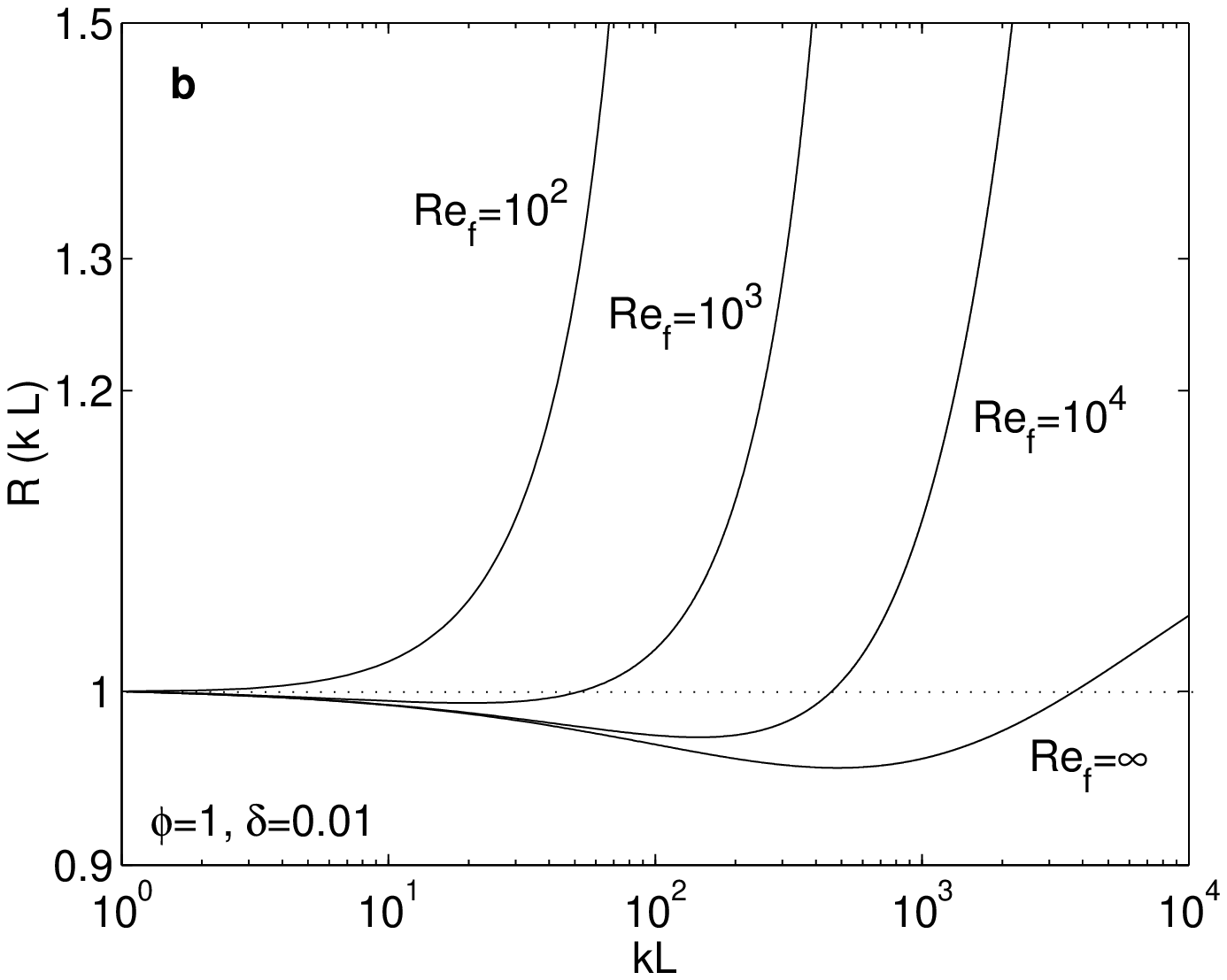} \\ %%%%%%%%%%%%%%%
\epsfxsize=7.5cm \epsfbox{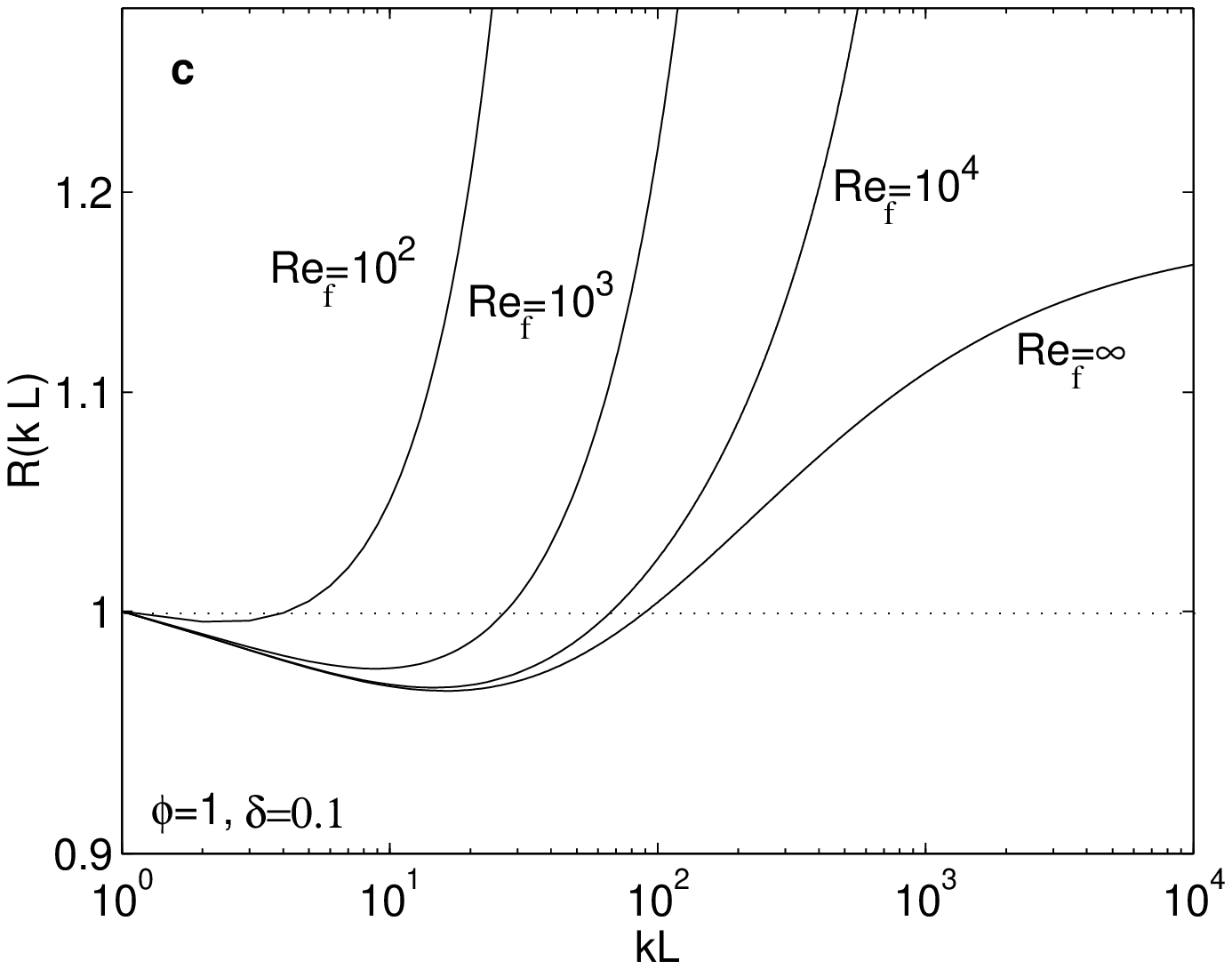} \\ %%%%%%%%%%%%%%%
\caption{\label{f:F4}                     %%%%%%%%%%%%%%%
The plots of the numerical results for $R(\kappa)$ ($\kappa=kL$) for
various $\Re\sb f$ (solid lines) for $\phi=1$ and $C=C_1=0.5$.  Panel
$a$: $\delta=10^{-3}$, dashed lines show analytical prediction,
Eq.~(\ref{eq:Rdelta0}) for $\delta=0$. Panels $b$ and $c$:
$\delta=0.01$ and $\delta=0.1$.}
\end{figure}
%%%%%%%%%%%%%%%%%%%%%%%%%%%%%%%%%%%

\subsection{Turbulence modification for finite $\Re$ }
\label{ss:modif-Re}

In the previous section we discussed the mechanism of turbulent
enhancement in the inertial interval caused by the density
variation in the energy cascade processes. There is one more
 mechanism of the turbulent enhancement, near the viscous subrange, that may be even more important at moderate $\Re$.
This effect is  due to the renormalization of the kinematic
viscosity in suspensions,  $\nu \Rightarrow \nu \sb{eff}(k)$,
\REF{nu-eff}, caused again by the density variation. Since $\rho
\sb{eff}(k )$ near the viscous cutoff, $k\sim 1/\eta$,  is larger
than $ \rho\sb f$ ( and consequently $\nu \sb{eff}(1/\eta)< \nu$), the extent of the
inertial interval in suspension is therefore larger than that in the
particle free case for the same energy pumping to the system. Within
our model this effect may be  described for very small particles
with a response time smaller than the turnover time of
$\eta$-eddies. In this case the effective density is $k$-independent
in the inertial subrange, $\rho_\kappa=1$, $\ve_\kappa$ and
$\ve^{(0)}_\kappa$ (for the particle free case) are given by Eqs.~\Ref{sol1}
and \Ref{sol0}. Thus \REF{ratio} yields
\begin{equation}\label{eq:Rdelta0}
R(\kappa)= \Big[1+\frac{C_1}{4\Re\sb
s}\left(1-\kappa^{4/3}\right)\Big]^2\Big /
\Big[1+\frac{C_1}{4\Re\sb f}\left(1-\kappa^{4/3}\right)\Big]^2  .
\end{equation}
Plots of $R(\kappa)$ for different $\Re\sb f$ are shown in
Fig.~\ref{f:F4}$a$ by dashed lines together with the numerical
results for a quite small $\delta=10^{-3}$, solid lines. The numerical
results for $\delta=0.01$ and $\delta=0.1$ are shown in
Figs.~~\ref{f:F4}$b$ and ~\ref{f:F4}$c$. With  $\Re\sb f$ growing
above $10^{6}-10^{8}$ we return back to the situation in the
inertial interval, described above, see Fig.~\ref{f:F3}. For the
comparison we show  the plots for $\Re\sb f\to \infty$ in
Fig.~\ref{f:F4}.

For very small $\delta$  the effect of particles on the turbulent
statistics in the inertial interval is negligible; as an
illustration  see Fig.~\ref{f:F4}$a$ for $\delta=10^{-3}$. In this
case there is only the viscous range enhancement. Clearly, with
decreasing  $\Re\sb f$ this effect is more pronounced. For the moderate
values of $\delta$,  there is a turbulence suppression in the
beginning of the inertial interval, which turns into a turbulence
enhancement  in the bulk of the inertial interval, see, e.g.
Fig.~\ref{f:F3}$b$ for $\delta=0.01$ and the line marked $\Re\sb f=\infty$
in Fig.~\ref{f:F4}$b$. One sees that already for $\Re\sb f=10^4$
the energy enhancement increases. This enlargement becomes more
and more pronounced for even smaller $\Re\sb f$. For $\Re\sb f=10^2$  the
turbulence suppression in the beginning of the inertial
interval is negligible. Further development of these  tendencies
is illustrated in Fig.~\ref{f:F4}$c$ for $\delta=0.1$
%%%%%%%%%%%%%%%%%%%%%%%%%%%%%%%%%%%
\subsection{Brief comparison with DNS} \label{ss:comp}
In order to  get an analytical description of  the main physical
mechanisms of the particle effect on turbulence  we used in this
paper as simple as possible approximations, which nevertheless
preserve the basic physics of the problem. In particular, we have
used the differential approximation of the energy flux term,
\REF{flux1} with local in $k$-space closure procedure, which gives
a reasonable approximation in the extended inertial intervals of several
decades. However,  in the direct numerical simulations of
turbulence in suspensions, e.g. in Ref.~\cite{98BSS}, there is
almost no inertial interval, definitely smaller than one decade.
Therefore the detailed comparison of our simple theoretical
picture with DNS may be only qualitative.

\begin{figure}
\epsfxsize=8.5cm \epsfbox{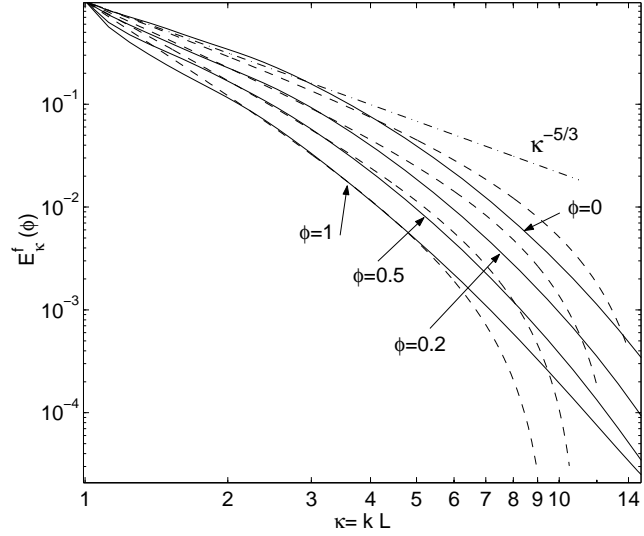} \caption{\label{f:exp}
Log-Log plots of turbulent kinetic energy spectrum $E\sp
f_\kappa(\phi)$ taken from \cite{98BSS} for $\phi=0, 0.2, 0.5$ and
1  with $\delta=1.65$ (solid lines), and numerical solution of
\REF{bud3}  for the same values of $\phi$ and  $\delta$ with
$\Re\sp{eff}\sb f=40$ and  $C=13$.}
\end{figure}

For such a comparison with DNS by Boivin, Simonin and
Squires~\cite{98BSS}) we re-plotted  in Fig.~\ref{f:exp} their
Fig.~5b for the kinetic energy spectra $E\sb s(k,\phi)$ of
suspensions in the log-log coordinates (solid lines). The solid
line, labelled  by $\phi=0$,  describes the particle free case, in
which the energy spectrum in the inertial interval should be scale
invariant. The K41 dependence is shown in Fig.~\ref{f:exp} by a
dash-dotted line, labelled by $\kappa^{-5/3}$. One sees that only
the first half of the decade may be considered as the inertial
interval. The viscous corrections to this dependence may be
accounted for with the help of \REF{sol0}. Using also
Eqs.~\Ref{Re-eff} and  \Ref{Efn}, one gets:
\begin{eqnarray}\label{eq:sol10}
E\sp f_\kappa= \kappa^{-5/3}\left[1 +\frac{1}{\Re\sb f\sp
{eff}}(1-\kappa^{4/3}) \right]^2\ .
\end{eqnarray}
With an appropriate value of $\Re\sp{eff}\sb f$ this equation
reasonably approximates the numerical data almost in the whole
decade of $\kappa$, in which   $E\sp f_\kappa$ decays more than
three orders of magnitude, see dashed line $\phi=0$. The chosen value
$\Re\sp{eff}\sb f=40$ agrees with parameters given in
Ref.~\cite{98BSS} with an acceptable value of the closure
parameter $C_1$, which enters in the definition~(\ref{eq:Re-eff})
for  $\Re\sp{eff}\sb f$. With $C=13$ the numerical solutions of
\REF{bud3} approximate well all  the DNS energy spectra $E_\kappa
\sp f(\phi)$ with $\phi=0.2,\, 0.5,$ and $1$ in the region,
bounded from above by some  value of $\kappa$ referred to as
$\kappa\sb {max}$. In this region the spectra decrease from unity
(at $\kappa=1$) to some values, smaller than $10^{-3}$. The value
of $\kappa\sb {max}$ decreases from $\kappa\sb {max}\approx 14$
for $\phi=0$ spectrum to $\kappa\sb {max}\approx 7$ for the spectrum
with $\phi=1$.

For $\kappa> \kappa\sb {max}$ the solutions of  \REF{bud3} give
too small values of the turbulent energy. As already discussed,
this is due to the differential approximation for the energy flux,
which is absolutely not realistic in the viscous subrange.
Clearly, the larger the value of $\phi$, the more energy is
dissipated by the fluid-particle friction, diminishing the energy
flux at the end of the inertial interval. Consequently, the
Kolmogorov microscale $\eta=\nu^{3/4}/\varepsilon^{1/3}$
increases.  Since $\kappa\sb {max} \propto 1/\eta$, it shifts
toward smaller values.

One observes also some deviations of the DNS data and our
numerical solutions in the energy containing region $\kappa\sim 1$.
This is again related to the differential approximation for the
energy flux. To improve the description of the particle effect on
the turbulent statistics a better approximation for the energy
transfer term is required. This calls for the  more elaborated
closure procedures, based on a proper analysis of the triad
interactions.

%%%%%%%%%%%%%%%%%%%%%%%%
%%%%%%%%%%%%%%%%%%%%%%%%%

\S{sum}{Summary}
 \textbullet~In this paper we propose a  \it{one-fluid} dynamical
model of turbulently flowing dilute suspensions which differs from the
usual  Navier-Stokes equation in two aspects:

\begin{enumerate}
\item
Instead of fluid density, $\rho\sb f$, our model involves a
$k$ dependent \it{effective density} $\rho\sb {eff}(k)$ which
varies between $\rho\sb f$ (for large $k$) and the mean density of
suspension $\rho\sb s=\rho\sb f(1+\phi)$ (for small $k$);

\item The model equation includes an additional damping term
$\propto \gamma\sb p(k)$ which describes the fluid-particle
viscous friction.

\end{enumerate}
\textbullet~Our model may be considered as a \it{mean-field
approximation} in which one uses a dynamical equation of motion
with ``effective'' coefficients which depend on the statistics of
the resulting  stochastic solutions. In our case $\rho\sb
{eff}(k)$ and $\gamma\sb p(k)$ are determined by the eddy turnover
frequency $\gamma(k)$ which, in its turn, depends on the resulting
energy distribution in the system.

\textbullet~Our model is based on the same set of assumptions
(applicability of the Stokes law for the fluid-particle friction
and space homogeneity of the particle distribution) as widely used
in \it{two-fluid} models for suspensions. We believe that the \it{one-fluid}
description of turbulent suspensions contains the same physics as
the essentially more complicated \it{two-fluid} models. Our
feeling is that a possible minor difference in the level of
accuracy between these two models is beyond a current level of
understanding of the problem and is definitely  smaller than the
``absolute'' accuracy of each model itself.

\textbullet~In order to keep the description of the problem as
simple and transparent as possible  we used in this paper a
closure procedure based on the Kolmogorov-41 dimensional reasoning
with an additional simplification --- the differential form of the
energy transfer term in which the energy flux $\ve(k)$ is
evaluated \it{locally in  $k$-space}, via the spectrum $\C
E(k)$ taken at the \it{same wave-number} $k$. This allows us to
derive the quite simple ordinary differential equation for the energy
budget in the system \Ref{bud3}.

\textbullet~As a reward, our budget  \REF{bud3} allows an
effective analytical analysis in various important limiting cases,
i.e.:
\begin{enumerate}
\item In the particle free case, see \REF{sol0}; %%%%%%%
\item For the micro-particles case ($\delta < \Re^{-3/4}$), see \REF{sol1}; %%%%%%%%%%
\item For the first decades of the inertial interval (in the case
 $\delta < 1$)   or in the whole inertial interval (if $C<1/4$),
see  Eqs.~\Ref{ve0} and \Ref{gamma0}; %%%%%%%%%%
\item For any reasonable values  of parameters at hand, see
Eqs.~\Ref{ve1} and \Ref{gamma1}, involving one-dimensional
integration.
\end{enumerate}
In the general case the budget equation \Ref{bud3} may be easily
solved numerically.

\textbullet~We derived the analytical expression \Ref{Rkappa} for
the dimensionless ratio $R_0(kL)$, which  describes the energy
suppression and enhancement  in the \emph{inertial interval of
scales}.

 \textbullet~In Sec.~\ref{ss:modif-Re}
we described the additional  "viscous" mechanism of the turbulence
suppression and  enhancement, caused by the particle effect on the
\emph{extent of the inertial interval}:
\begin{enumerate}
\item The decrease  of the effective kinematic viscosity in
suspensions (due to the increase in the effective density for
small scale motions) \emph{ elongate the inertial interval}.

\item The fluid-particle friction causes a decrease of the energy
flux at  the viscous end of the inertial interval and hence \emph{
shorten the inertial interval}.
\end{enumerate}
The winner of this competition depends mainly on the value of
$\delta$, see, e.g. Fig.~\ref{f:F4}.

The complicated interplay of the inertial-range and the
viscous-range mechanisms of the suppression and the enhancement of
the turbulent activity in suspensions is the main topic of
Sects.~\ref{s:prelim} and \ref{s:modification}.

\textbullet~Our model successfully correlates observed features of
numerical simulations. These features are the following:

\begin{enumerate}

\item For a suspension with particles with a response time much
larger than the Kolmogorov time the main effect of the particles
is suppression of the turbulence energy of fluid eddies of all
sizes (at the same energy input as for the particle-free case).
See for instance Fig.~\ref{f:exp}, where a comparison with the
DNS-results of Boivin, Simonin and Squires~\cite{98BSS} is shown.

\item  For a suspension with particles with a response time
comparable to or smaller than the Kolmogorov time the Kolmogorov
length scale of the fluid eddies will decrease and the turbulence
energy of eddies of (nearly) all sizes increases (at the same
energy input as for the particle-free case). This result was also
reported by Druzhinin~\cite{01Dru}, who carried out
DNS-simulations for the case of micro-particles.

\item  For a suspension with particles with a response time in
between the two limiting cases mentioned above the energy of the
larger fluid eddies is suppressed whereas the energy of the
smaller eddies is enhanced. The cross-over between suppression and
enhancement depends on the ratio of the particle response time and
the Kolmogorov time. The strength of the effect depends on the
mass loading. This is in agreement with the DNS-results of
Sundaram and Collins~\cite{99SC}.

\end{enumerate}

The more detailed comparison of our approach to turbulent
suspensions with the physical and numerical experiments requires:
\begin{enumerate}
\item From DNS side more detailed  analysis of joint statistics of
the velocity field of the particle and the carried fluid;

\item From the theoretical side  an application of the more
advanced \it{non-local} closure procedures, explicitly accounting
for  the triad interactions.
\end{enumerate}

\textbullet~An additional advantage of our one-fluid approach is
that one can use standard and well developed closures from
analytical theory of one-phase turbulence. This fact and the relative
simplicity and physical transparency of the one-fluid model
equations may essentially help in the further progress towards a
theory of turbulent suspensions for more realistic cases with
space inhomogeneities, gravitational settling, \etc

\subsection*{Acknowledgements}
We thank T. Elperin, N. Kleeorin, and I. Rogachevskii for helpful
discussions, which contributed to this paper.  This work has been
partially supported by the Israel Science Foundation and the
Nitherlands Foundation of Applied Sciences. Two
of us (V.L and A.P)  acknowledge the hospitality at the
Burgerscentrum, Delft, The Netherlands.

\Twocol
\end{document}